\documentclass[onecolumn]{IEEEtran}
\usepackage[T1]{fontenc}
\usepackage[utf8]{inputenc}
\usepackage{varwidth}
\usepackage{amsmath}
\usepackage{amsthm}
\usepackage{graphicx}

\makeatletter

\providecommand{\tabularnewline}{\\}
\newenvironment{cellvarwidth}[1][t]
    {\begin{varwidth}[#1]{\linewidth}}
    {\@finalstrut\@arstrutbox\end{varwidth}}

\theoremstyle{plain}
\newtheorem{thm}{\protect\theoremname}
\theoremstyle{definition}
\newtheorem{defn}[thm]{\protect\definitionname}
\theoremstyle{definition}
\newtheorem{example}[thm]{\protect\examplename}
\theoremstyle{remark}
\newtheorem{rem}[thm]{\protect\remarkname}
\theoremstyle{plain}
\newtheorem{lem}[thm]{\protect\lemmaname}
\theoremstyle{plain}
\newtheorem{cor}[thm]{\protect\corollaryname}

\providecommand{\tabularnewline}{\\}

\providecommand{\corollaryname}{Corollary}
\providecommand{\definitionname}{Definition}
\providecommand{\lemmaname}{Lemma}
\providecommand{\remarkname}{Remark}
\providecommand{\theoremname}{Theorem}

\makeatother

\usepackage{babel}

\addto\captionsamerican{\renewcommand{\corollaryname}{Corollary}}
\addto\captionsamerican{\renewcommand{\definitionname}{Definition}}
\addto\captionsamerican{\renewcommand{\examplename}{Example}}
\addto\captionsamerican{\renewcommand{\lemmaname}{Lemma}}
\addto\captionsamerican{\renewcommand{\remarkname}{Remark}}
\addto\captionsamerican{\renewcommand{\theoremname}{Theorem}}
\addto\captionscanadian{\renewcommand{\corollaryname}{Corollary}}
\addto\captionscanadian{\renewcommand{\definitionname}{Definition}}
\addto\captionscanadian{\renewcommand{\examplename}{Example}}
\addto\captionscanadian{\renewcommand{\lemmaname}{Lemma}}
\addto\captionscanadian{\renewcommand{\remarkname}{Remark}}
\addto\captionscanadian{\renewcommand{\theoremname}{Theorem}}
\addto\captionsenglish{\renewcommand{\corollaryname}{Corollary}}
\addto\captionsenglish{\renewcommand{\definitionname}{Definition}}
\addto\captionsenglish{\renewcommand{\examplename}{Example}}
\addto\captionsenglish{\renewcommand{\lemmaname}{Lemma}}
\addto\captionsenglish{\renewcommand{\remarkname}{Remark}}
\addto\captionsenglish{\renewcommand{\theoremname}{Theorem}}
\providecommand{\corollaryname}{Corollary}
\providecommand{\definitionname}{Definition}
\providecommand{\examplename}{Example}
\providecommand{\lemmaname}{Lemma}
\providecommand{\remarkname}{Remark}
\providecommand{\theoremname}{Theorem}

\begin{document}
\title{Energy-Time Optimal Decoding Using Expander-Graph LDPC Codes and Mesh
Network Architecture}
\author{Christopher Blake\\
Great Bay University\\
\texttt{\small christopher.blake@gbu.edu.cn} }
\maketitle
\begin{abstract}
Within the Thompson VLSI model, the first energy optimal decoding
schemes (up to polylogarithmic factors) that also reach clock cycle
lower bounds are constructed. Let $N$ be block length. Then, there
is an exponential error scaling function and a binary symmetric channel,
such that for all $f(N)$ that scales to zero more slowly than this,
there is a decoding scheme for this channel with error probability
that scales as $O(f(N))$ that has number of clock cycles $T\le O(\sqrt{-\log(f(N)})\text{polylog}(N))$
and energy $E\le O(N\sqrt{-\log(f(N))}\text{polylog}(N))$. This matches
fundamental lower bounds up to polylogarithmic factors. The technique
uses expander-graph based LDPC codes, a mesh network architecture,
and a parallelization technique. The key technical result needed for
this is a lemma in graph theory. This lemma is used to show that communication
graphs of bounded node degree can be routed through two- and three-dimensional
mesh networks without overflowing the memory of any mesh node. Upper
bounds are also extended to three dimensions.
\end{abstract}

\section{Introduction}

\IEEEPARstart {M}{any} computational algorithms, including those for error control coding,
belief propagation, and machine learning, can be characterized as
parallel computations on graphs. Error control coding is implemented
on cellphones and distributed devices, and thus requires fast and
efficient computation that consumes very low energy. On the other
hand, modern artificial intelligence and machine learning algorithms
require extraordinary amounts of energy \cite{delYeyDeepLearningEstimatingEnergy}.
These requirements -- for both hyper--energy-efficient error control
coding for small, battery-powered devices and for optimal energy use
in massive data center architectures -- motivate this paper. Our
main focus is on scaling rules, and in particular scaling rule \emph{upper
bounds}, for a large number of computational problems defined on graphs.
We apply this to the problem of error control coding, where we construct
the first energy-time optimal error control decoding schemes.

In particular we consider the \emph{mesh network}, a natural design
architecture first theoretically analyzed by Thompson \cite{ThompsonThesis,Thompson}.
Since three-dimensional circuit layout techniques exist \cite{dhananhayMonolithic3d},
we also generalize the approach to three dimensions. A mesh network
is a two- or three-dimensional grid of Turing-complete processors,
or nodes, in which each node can communicate information to its adjacent
nodes. 

This work is a generalization of Chapter 6 in the thesis of Blake
\cite{blakePhDThesis}, which has not yet been published in a peer-reviewed
journal. The main result of this work involves Lemma \ref{lem:LemmaScalingLemma}.
In this Lemma, we consider a sequence of graphs with $N$ vertices
and constant maximum node degree. We consider permuting the vertices
of this graph onto a set of $N$ slots, where two disjoint subsets
of these slots are considered distinguished. Subject to some conditions
on the sizes of those slots, we show that this random permutation
results in at most $O(\log N)$ edges connecting the two distinguished
subsets, with high probability.

The idea is that the edges of the graph can correspond to messages
to be sent between computational nodes in the parallel message-passing
step of distributed algorithms. The lemma then implies that there
exists a way to lay out a graph on a mesh network that allows for
a message-passing procedure without exceeding the computational or
memory requirements of each mesh node.

We apply this result directly to energy consumption of any algorithm
defined on a graph. This includes matrix-vector multiplication, LDPC
decoding on a Tanner graph, graph-based neural networks with constant
node degree, and any other parallel-computation algorithm that can
be defined on a graph. The key component is that the graph associated
with each of these computations has bounded maximum node-degree. We
show that for each of these types of computations, the communication
step can be accomplished in parallel with a straightforward parallel
message-passing protocol. For the two-dimensional case, the natural
message passing protocol that we propose is called the up-down left-right
protocol, and for three dimensions, we call it the up-down left-right
back-forth protocol.

This result is a key ingredient in the first construction of energy-time
optimal decoding schemes. These are schemes that reach energy consumption
lower bounds, and also simultaneously reach the time lower bounds
for such energy optimal schemes, up to polylogarithmic factors. The
approach uses the expander-graph based LDPC codes of Burshtein and
Miller \cite{burshteinExpanderGraph}. Notably, the connection to
the Expander Graph paper was suggested by ChatGPT Pro when proofreading
an earlier version of this paper. We show how a decoding scheme with
an exponentially vanishing probability of error that is also energy
and time optimal (up to polylogarithmic factors) can be used to construct
any energy-time optimal coding scheme with error scaling $f(N)$ using
a straightforward parallelization argument, for any $f(N)$ that is
slower than this exponential scaling.

\textbf{Note: }In this paper we use big-oh and big-theta notation
introduced by Bachmann and Landau \cite{bachmannNotation,landauNotation}.
When we say $g(x)=\Theta(f(x))$, we mean that there are two positive
constants $b$ and $c$ such that, for sufficiently large $x$, $bf(x)\le g(x)\le cf(x)$.
The notation $g(x)=O(f(x))$ is for upper bounds, and means there
is a positive constant $c$ such that, for sufficiently large $x$,
$g(x)\le cf(x)$. Finally, $g(x)=\Omega(f(x))$ is for lower bounds,
which means there is a positive constant $c$ such that, for sufficiently
large $x$, $g(x)\ge cf(x)$. The notation $g(x)=o(f(x))$ is used
to denote strictly loose upper bounds. It means that $\lim_{x\rightarrow\infty}g(x)/f(x)=0$. 

\section{Paper outline}

We first introduce some prior work in Section \ref{sec:Prior-work}.
We then introduce a new model of computation called the Three-Dimensional
Thompson Model in Section \ref{sec:Three-dimensional-Thompson-Model}.
This is a natural generalization of the two-dimensional Thompson model,
and within this model, we argue that a three-dimensional mesh network
can be instantiated, which we introduce in Section \ref{sec:meshNetworkSection}.
Then, in Section \ref{sec:CommunicationGraphDefinition}, we define
the communication graph of an algorithm, and discuss in general how
the messages corresponding to the communication graph can be passed
along on the mesh network architecture. In particular, we illuminate
the issue of \emph{routing conflicts}, which occur when too many messages
are routed through a single mesh node during the communication step,
overflowing the memory of the node. In Section \ref{sec:MessagePassingProtocolDescription}
we then describe two specific message passing protocols, the up-down
left-right protocol for two dimensions, and the up-down left-right
back-forth protocol for three dimensions, which are the main subject
of our analysis for the rest of the paper. In Section \ref{sec:mainTheoremOutline}
we introduce our main theorem, and in Section \ref{sec:mainTechnicalLemma}
we introduce and prove the main technical contribution of our paper:
the Graph Placement Lemma. Sections \ref{sec:twoDimensionalMessagePassingResult}
and \ref{sec:Three-Dimensional-Message-Passing} then apply this Lemma
to the two- and three-dimensional cases, which prove our primary theorems:
that the message passing protocols we introduce can implement graphs
of bounded node degree in both two- and three-dimensional mesh networks
without encountering routing conflicts that overflow the memory of
a mesh node. In Sections \ref{sec:LDPCScalingSection} and \ref{sec:SparseMatrixVector},
we apply these results to the LDPC decoding problem and sparse matrix-vector
multiplication problem, respectively. In Section \ref{sec:parallelConstruction}
we show how using an exponential error probability coding scheme that
is energy-time optimal can be then used to construct an energy-time
optimal coding scheme for any sub-exponential error probability. Finally,
the construction of such a scheme using expander graph based LDPC
codes is presented in Section \ref{sec:ExpanderGraphEnergyTimeOptimality}.

\section{Prior work}\label{sec:Prior-work}

This work stems from results in the thesis of Blake \cite{blakePhDThesis}.
In this thesis, the author considers the two-dimensional mesh network
case. The results in this paper generalize this result to three dimensions.
Because the two-dimensional result has not yet been published in a
peer reviewed paper, we also present those results. The key idea of
this paper is a generalization of Lemma 25 of the thesis, which we
call the Graph Placement Lemma\emph{. }We also correct the thesis
in a minor way: specifically, we argue that the routing costs incur
an extra $\log\log N$ factor when there is only one processor in
each mesh node. 

The graph theoretic result is then applied to computation on a \emph{mesh
network}, a computational architecture introduced by Thompson \cite{ThompsonThesis,Thompson}.
The key results of this approach were energy consumption scaling rule
\emph{lower bounds} for the sorting and discrete Fourier transform
problems; Thompson then showed that these lower bounds could be reached
up to polylogarithmic factors. The lower bound technique has also
been extended to integer multiplication \cite{AblesonInformationTransferVLSIMultiplication}.

On the error control coding question, \cite{groverFundamental} provided
some lower bounds on energy consumption of decoders for the Thompson
model. However, the authors did not provide any \emph{upper bounds.
}Grover, in \cite{GroverInfoFriction}, introduced the \emph{information
friction }model, in which energy consumption is proportional to the
number of bits communicated in a circuit multiplied by the distance.
This approach was extended to three dimensions in \cite{blakeKschischangEnergyLatencyReliabilityTradeoffs}.
Recently, \cite{gianinazzi2023spatialcomputermodelenergyefficient}
introduced a similar parallel model of computation, in which, like
the Grover information friction model, energy is proportional to the
total distance travelled by the bits. 

Recently, authors have introduced a model similar to the Thompson
model \cite{jinWireDelay}. In this approach, instead of considering
energy consumption, the authors consider the \emph{time delay }of
passing messages along wires. In the model we consider here, however,
the time delay to propagate along a wire is considered negligible,
and the time delay of passing through a logic gate dominates. As the
authors point out, however, as we push to the physical extremes of
computation, this assumption may no longer hold; nevertheless we do
not consider these wire delays in this paper. Moreover, these authors
consider \emph{mildly} three-dimensional circuits, with up to a constant
number of layers stretching to the third dimension. However, since
we are concerned with \emph{fundamental }physical limits of computation
here, and since multiple different circuits themselves can be placed
physically on racks, we place no limitation on the number of vertical
layers in our system. 

The key ingredient to finally achieve energy-time optimal coding schemes
comes from \cite{burshteinExpanderGraph}, which showed that LDPC
coding schemes exist with bounded maximum node degree and a logarithmic
number of iterations that can correct some constant fraction of errors.

\section{Three-Dimensional Thompson Model}\label{sec:Three-dimensional-Thompson-Model}

\subsection{The Three-Dimensional Thompson Model Description}

Please refer to \cite{ThompsonThesis} or \cite{blakePhDThesis} for
a description of the Two-Dimensional Thompson model. We generalize
this model to three dimensions here, and call this the Three-Dimensional
Thompson Model. Note that this is a description of a three-dimensional
circuit \emph{in general}. To avoid confusion, we emphasize that only
in the \emph{next} Section will we describe a specific instance of
a three-dimensional circuit called the \emph{mesh network, }which
is the main focus of this paper. Even though they are both cubes,
do not confuse a grid cube described in \emph{this} section with a
mesh-node, which we will introduce in the next section. A mesh node
will be much bigger and include memory and processing elements.
\begin{enumerate}
\item A computation is done on a grid of cubes, with sides of unit length.
They are also referred to as nodes, which should not be confused with
the mesh nodes in the mesh network, or the computational node of a
graph. Which type of ``node'' is referred to shall be clear from context.
\item The computation is divided into distinct clock cycles.
\item Each cube can be a computational grid cube (computational node) or
a wire grid cube (wire node).
\item Wire grid cubes can transport binary information (either a $0$ or
$1$). Generally, a wire grid cube has one face acting as the \emph{in-face,
}which can take in input from the adjacent grid cubes. Any subset
of the remaining faces of the grid cube can be designated as \emph{out-faces,
}which route the bit injected into the in-face to the adjacent grid
cubes, which can be either a computational grid cube or a wire grid
cube.
\item Adjacent and connected wire grid cubes form a directed graph form
of a rooted tree, which in graph theory is also called an \emph{arborescence}
\cite{jinWireDelay}. The root of the tree must be connected to the
output of a computational node, and the leaves of this tree must be
connected to an input of a computational node. In a single clock cycle,
the message injected into the root of this tree at the beginning of
the clock cycle is propagated across the entire wire to the leaves.
In other words, this tree behaves just like a normal node in a digital
circuit, with a message propagated throughout the entire interconnected
wire in a single clock cycle, until it reaches all the computational
elements to which it is connected.
\item Computational grid cubes can take in input from any of their six faces,
which may be wires or other computational nodes. They can output computations
out of any of their six output faces. A face of the computational
grid cube may also remain unused, and have no inputs or outputs. In
a single clock cycle, the computational nodes take in the input messages
from their input faces, compute a fixed boolean function of these
inputs, and then output to all the output faces.
\item All input bits are injected into input nodes at the beginning of the
computation. That is, we do not consider serialization of the inputs
in this paper.
\item Computational nodes can compute any boolean function mapping their
inputs to their outputs.
\item A computational node takes exactly one clock cycle to compute its
output, at which point its output wires are potentially switched from
a $0$ to a $1$ or vice versa. The total number of clock cycles required
to do the computation is time $T$. 
\item \textbf{Key Idea: }The energy of a computation in each clock cycle
is proportional to the number of grid cubes that are \emph{active}
in that clock cycle. A wire grid cube is considered active if it switches
from a $0$ or $1$ during the course of the computation; a computational
grid cube is considered active if it processes any input that changed
from the previous clock cycle. We call $q$ the \emph{switching activity
factor, }which is the average fraction of grid cubes that are active
per clock cycle. We let $V$ be the total number of grid cubes occupied
in the circuit either by a wire or a computational node (a quantity
we call the \emph{volume}). Thus energy consumption is equal to $E=qVT.$
\end{enumerate}
Just as in the two-dimensional model, the model flows from the idea
that a circuit acts as a capacitor, and the energy to charge and discharge
the circuit is equal to $\frac{1}{2}CV_{s}^{2},$ where $V_{s}$ is
the voltage used to run the circuit, and $C$ is the capacitance.
The capacitance of a wire is roughly proportional to its length, and
the total length of the wires is roughly proportional to the number
of occupied grid cubes. Hence, capacitance, and thus energy, is roughly
proportional to the volume of the circuit.

Note that we allow for the input nodes of the computation to be \emph{anywhere
}within the circuit, including its interior. This may conflict with
physical circuit designs in which inputs are injected into the perimeter
of the circuit. However, we use this model because it is physically
general, and it allows us to model computations where the inputs of
the computation are generated within the circuit itself, and thus
may be in the interior of the circuit at the beginning of the computation.

Obviously, since the distance any bit follows in such a circuit must
at least equal the Euclidean distance between source and destination,
lower bounds derived within the three-dimensional information-friction
model \cite{blakeKschischangEnergyLatencyReliabilityTradeoffs} would
seem to apply as lower bounds within this more physical Thompson model.
The only caveat is that for some instances of the computation, the
same bit value may be communicated multiple times within the Thompson
model and not ``switch'' any circuit elements. These bit communications
would cost something in the information friction model, but might
not cost anything within the Thompson model. 

\subsection{Logic gate fan-in and fan-out}

In our model, we limit the fan-in of the logic gates due to geometric
considerations: each node is a square or a cube, with a limited number
of edges and faces. This assumption is not important for lower bounds,
which have been the focus of prior work. However, in considering upper
bounds, it becomes more important to consider fan-in and fan-out.
In the model, \emph{time }is dominated by the number of logic gates
through which the execution must flow. This is meant to capture the
idea that in many real circuits, the depth of logical operations dominates
computational time. However, some simple logic gates may be implementable
with larger fan-in, while still allowing for (roughly) constant-time
execution. Computation that allows for unbounded fan-in AND and OR
gates is captured in the computational complexity class $AC^{0}$
(See \cite{HastadThesis} for a discussion of small depth circuits
and a classic lower bound result for the computation of parity). Our
model only allows for bounded fan-in logic gates. Also, our routing
computations will be on addresses of size $O(\log N)$. Rudimentary
arguments requiring outputs to ``see'' all the inputs mean that many
computations, including arithmetic computations needed in our routing
algorithms, require depths logarithmic in input size. This adds a
$\log\log(N)$ factor to our complexity upper bounds. This factor
was neglected in the thesis upon which this work was based \cite{blakePhDThesis},
which we correct in this paper. Unlimited fan-in logic gates could
potentially remedy this issue, which we do not consider herein. 

\section{The Mesh Network}\label{sec:meshNetworkSection}

\subsection{Description of Mesh Network in Two and Three Dimensions}

In this paper, we consider two types of mesh networks, the two-dimensional
mesh network and the three-dimensional mesh network.

In the two-dimensional mesh network, the circuit is a $\sqrt{N}\times\sqrt{N}$
grid of square computational nodes, each with side length $\Theta(\log N)$,
and area $\Theta(\log^{2}N)$. In two dimensions, each node is connected
to up to $4$ other computational nodes. 

\begin{table}
\begin{tabular}{|c|c|c|c|}
\hline 
Mesh Network type & Side Length of computational node & \begin{cellvarwidth}[t]
\centering
Maximum number of bits communicated 

in one clock cycle to adjacent node
\end{cellvarwidth} & Area or volume of each mesh node\tabularnewline
\hline 
\hline 
2D & $\Theta(\log N)$ & $\Theta(\log N)$ & Area=$\Theta(\log^{2}N)$\tabularnewline
\hline 
3D & $\Theta(\log^{\frac{2}{3}}N)$ & $\Theta(\log^{\frac{4}{3}}N)$ & Volume=$\Theta(\log^{2}N)$\tabularnewline
\hline 
\end{tabular}\caption{Table representing the side length, number of bits that can be communicated
in one clock cycle to an adjacent node, and the area or volume of
each node. Note that side length is chosen so that the area/volume
of each node is $O(\log^{2}N)$. This allows for up to $O(\log N)$
message-address pairs to be routed through a node at a specific one-hop
communication step, where each address has size $O(\log N)$.}\label{tab:tableOfMeshNetworkProperties}

\end{table}

In the three-dimensional mesh network, the circuit is a $\sqrt[3]{N}\times\sqrt[3]{N}\times\sqrt[3]{N}$
grid of cubic computational nodes, each with side length $\Theta\left(\log^{\frac{2}{3}}N\right)$,
and each with volume $\Theta(\log^{2}(N))$. In the three-dimensional
mesh network, each node is connected to the (up to 6) adjacent nodes.

See Table \ref{tab:tableOfMeshNetworkProperties} for a summary of
the properties of the two- and three-dimensional mesh networks. Note
that the dimensions of the two- and three-dimensional mesh networks
are chosen so that the area/volume of each computational node is $\Theta(\log^{2}N)$,
which allows each node to hold up to $O(\log(N))$ addresses, each
of size $O(\log(N))$. 

Each mesh node is instantiated with a single Turing-complete processor.
Hence, we call the mesh architecture with one processor per mesh-node
the \emph{Single-Processor Node Mesh Network}.

The purpose of a mesh network is to enable message passing between
adjacent nodes, and also computation upon those messages that were
received. We will consider using this architecture to implement message-passing
algorithms that have a message-passing component and a computational
component.

\subsection{Implementation Detail and Assumptions}

The circuit that decides in which direction a message should be routed
must be able to handle messages of size $\log N$. Moreover, it must
be able to compare the $x$ and $y$ coordinates of the node's own
address with the destination address. This can be accomplished with
a subtraction circuit. This is achievable with a classic construction
of a carry-lookahead adder or other possible implementation, which
takes depth that is logarithmic in the input length \cite{Ofman1962OnTA}.
Note that since our input length is $O(\log(N))$, within our model
this will take depth $O(\log\log(N)).$ Such circuits would have width
$O(\log(N))$ and depth $O(\log\log(N))$, and thus area $O(\log N\log\log N)$,
and thus would fit into the node size of $O(\log^{2}N)$. 

We make an assumption about three-dimensional mesh networks, which
we call the \emph{3D-Circuit Folding Assumption}. We assume that processors,
memory, and routing wires can be squeezed into the three dimensional
grid cube without any overhead that changes asymptotic scaling. That
is, we assume the processor can take volume $O(\log N\log\log N)$,
the memory takes volume $O(\log^{2}N)$, and the total volume of all
of these components, including connecting wires, and input and output
routing wires, can be squeezed into the grid cube of side length $O(\log^{2/3}N)$.
The specific implementation details of such a construction are beyond
the scope of this paper; our focus is on understanding whether a communication
graph can be routed through the mesh network without overflowing memory
due to routing conflicts. Technically, however, the existence of such
a grid cube with such scaling properties is an unproven assumption,
which we leave as an area of future work.

\section{Communication graphs of an algorithm}\label{sec:CommunicationGraphDefinition}

\subsection{Communication graph definition}

We consider parallelized algorithms in general. Such computations
have computational nodes, each of which is capable of performing part
of the computation. At certain intervals as defined by the algorithm,
the computational nodes pass messages to other computational nodes
in a \emph{message-passing step. }

\begin{defn}
The \emph{communication graph} of an algorithm is the directed graph
of $N$ vertices. It represents the messages to be sent between computational
nodes during the computation.
\end{defn}

\begin{example}
In LDPC decoding, the communication graph is the Tanner graph of the
LDPC code.
\end{example}
\begin{example}
Consider computing the output of matrix-vector multiplication, where
$\boldsymbol{y=}\boldsymbol{Mx}$, in which \textbf{$\boldsymbol{M}$}
is a fixed and sparse $[n\times n]$ matrix, and the output $\boldsymbol{y}$
and input $\boldsymbol{x}$ are $[n\times1]$ vectors. A message passing
step for this computation involves the communication of input bits
to each of the $n$ nodes responsible for computing the output. Each
of those $n$ nodes computes a linear combination of its inputs. Thus,
the communication graph is that graph induced by the non-zero entries
of the matrix. See Figure \ref{fig:matrixVectorGraphDiagram} for
an illustrative example.
\end{example}

In general, a parallel computation can have multiple communication
steps, each corresponding to a different communication graph. For
our scaling laws, we consider a sequence of communication graphs in
which $N$ approaches infinity and where the maximum node degree is
bounded by a constant $d_{\max}.$

\subsection{Communicating on a mesh network}

For each vertex of a communication graph, we assign it to a computational
node of the mesh network.

\begin{defn}
The \emph{orientation} of a graph with $N$ vertices on a mesh network
of $N$ nodes is an assignment of the graph nodes to the nodes of
the mesh network.
\end{defn}
\begin{rem}
Technically speaking, as described, a mesh network must have a number
of computational nodes proportional to an integer squared or cubed,
but a graph does not necessarily have this many vertices. For our
discussion, we can just assume that the graph has additional isolated
vertices added to it so that the number of vertices is the appropriate
number. This is not a significant part of our derivation.
\end{rem}
Note that there are $N!$ possible orientations for a given graph.

In this setup, computation on a mesh network consists of communication
steps and computation steps. 
\begin{defn}
A \emph{communication step} is part of the computation process in
which the processor nodes send messages to other processor nodes in
parallel according to the communication graph. Since a processor node
can only send messages to adjacent nodes, to fully send all the messages
associated with the computation graph, the communication steps must
be divided into multiple \emph{one-hop communication steps}, or more
compactly, \emph{one-hop steps}. Before every such step, we assume
that each processor node contains message-address pairs in its memory
that are to be sent to another node in the circuit. In each such one-hop
step, each node must send all messages collected during the last one-hop
step to appropriate adjacent nodes, in a direction defined by the
selected protocol.
\end{defn}

The protocols we will consider are the up-down left-right protocol
for two dimensions, and the up-down left-right back-forth protocol
for three dimensions, which we describe in subsections that follow.

\begin{defn}
The \emph{computation step }consists of processing the messages received
after the communication step.
\end{defn}

\begin{defn}
A \emph{source node }is the source of a message. It corresponds to
a vertex of the communication graph with an outgoing edge.
\end{defn}
\begin{defn}
A \emph{destination node }is a node to which a source node is sending
one of its messages. It corresponds to a vertex in the communication
graph with an incoming edge.
\end{defn}
\begin{defn}
A \emph{routing node }is a node under consideration that must route
messages from source nodes to destination nodes.
\end{defn}
Note that routing nodes can be (and usually are) also source nodes
and destination nodes. However, if we fix a particular communication
graph and message-passing protocol, for any specific routing node
and any specific one-hop step, there is only a limited set of possible
source nodes and destination nodes for messages that could be routed
through this routing node. This is particularly true for the two message
passing protocols we analyze in this paper. The analysis of the size
of these sets is the main goal of Sections \ref{sec:twoDimensionalMessagePassingResult}
and \ref{sec:Three-Dimensional-Message-Passing}.
\begin{defn}
A \emph{routing conflict}, or more briefly a \emph{conflict, }occurs
when, during a communication step, multiple messages are routed through
a single mesh node at the same time. When
$\kappa$ messages are received by a single processor node after a
single one-hop step, such an event is called a $\kappa$-\emph{conflict}
and $\kappa$ is the \emph{size} of the conflict. See Figure \ref{fig:routingConflictDiagram}
for an example of a routing conflict of size $3$ at the third one-hop
step, when using the up-down left-right protocol which we will define
in the next section.
\end{defn}

We let $\kappa_{max}$ be the maximum number of message-address pairs
contained in the memory of any processor node during the execution
of the algorithm. By dividing the one-hop communication step into
$\kappa_{max}$ clock cycles, all the messages stored in the memory
of each processor can be communicated from one adjacent node to another
during each one-hop communication step, so long as the memory in any
of the computational nodes is not overflowed. 
\begin{rem}
Recall that the address part of each message has size $\Theta(\log N).$
Thus, so long as $O(\log(N))$ or fewer message-address pairs are
routed through any of the computational nodes at any one time, then
the maximum amount of memory needed to store messages being routed
is $O(\log^{2}N)$, which grows at the same rate as the size of the
memory of each mesh node. This will mean that there is no conflict
that overflows the memory of any computational node.
\end{rem}

\section{The Message Passing Protocols}\label{sec:MessagePassingProtocolDescription}

\subsection{Two Dimensions: Up-down left right protocol}\label{subsec:Two-Dimensions:-Up-down}

For the two-dimensional mesh network, we consider a very straightforward
protocol called the up-down left-right protocol.

\begin{figure}

\includegraphics[scale=0.5]{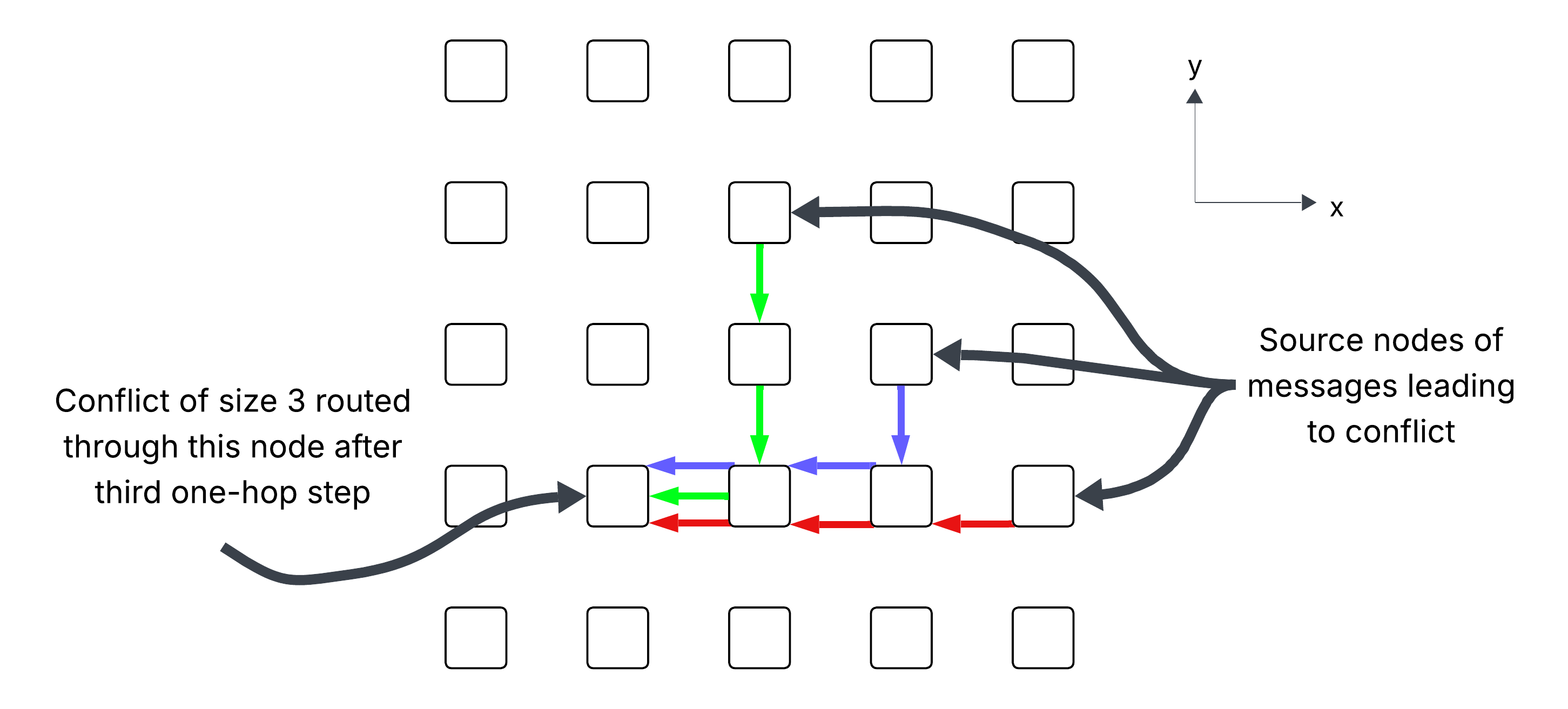}

\caption{Example of messages being routed through a two-dimensional mesh network,
using the up-down left-right protocol, resulting in a size $3$ conflict
at one-hop step $3$ at the indicated routing node. The messages start
at source nodes and each follows a path indicated by arrows of the
same colour. The destination nodes of these messages are not indicated
in the diagram. Nevertheless, the destination nodes must occur to
the left of the indicated routing node. }\label{fig:routingConflictDiagram}

\end{figure}

For this protocol, at the beginning of the communication step, each
computational node has a list of message-address pairs. To determine
in which direction a node should send a message (up, down, left, or
right), the processor node looks at the destination address. If the
destination node is not on the same y-coordinate, this node shall
send the message parallel to this axis in a direction closer to the
destination node. If the destination is on the same y-coordinate,
then the node shall look to the x-axis, and send left-right, parallel
to this axis, closer to the destination. In Figure \ref{fig:routingConflictDiagram}
we illustrate how this protocol works for three messages being sent
simultaneously.

\subsection{Three Dimensions: Up-down Left-right Back Forth Protocol}\label{subsec:3DUpdownLeftRightBackForthProtocolDefinition}

\begin{figure}

\includegraphics[scale=0.3]{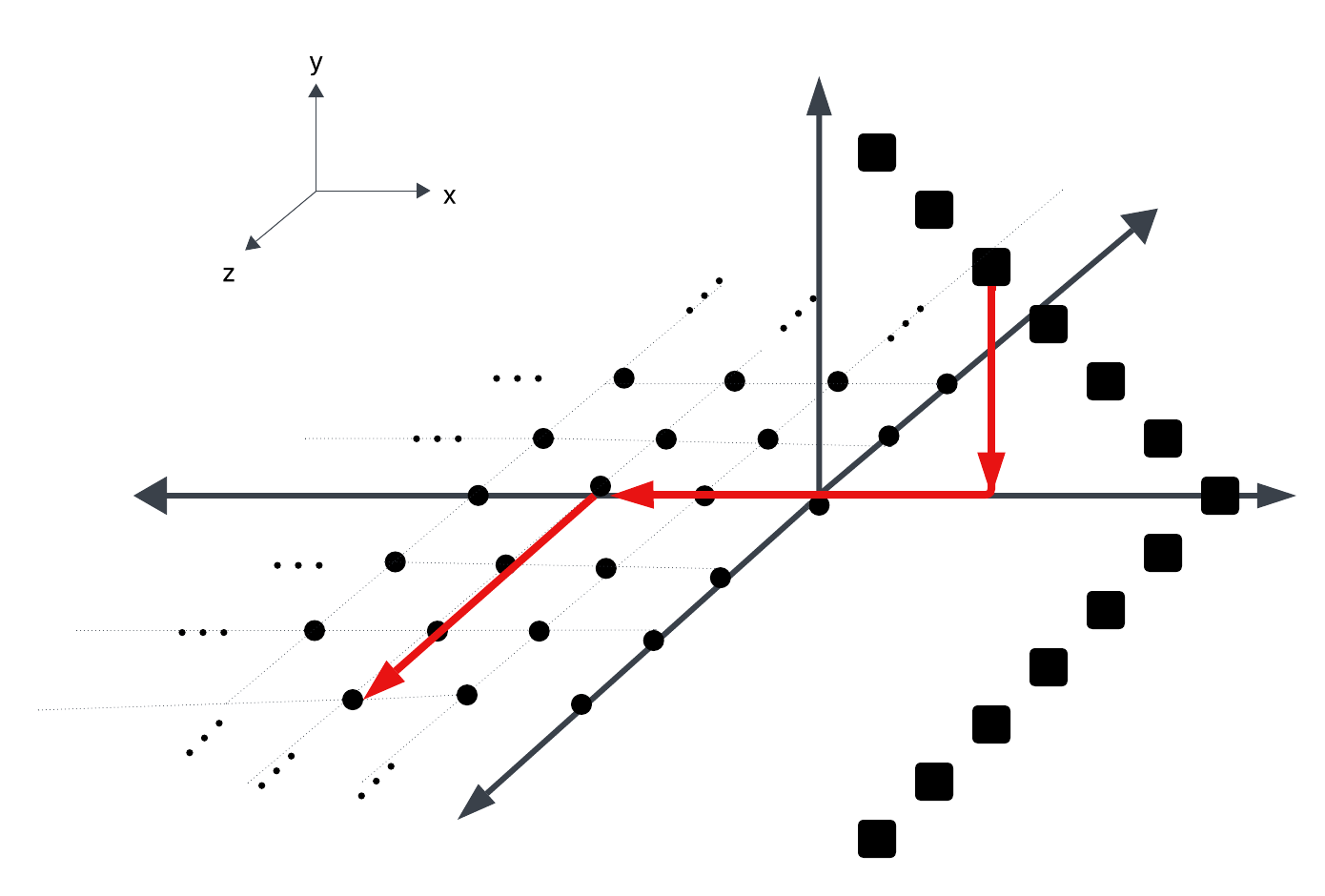}

\caption{Example of one path of the up-down left-right back-forth protocol.
The path is represented by a sequence of red arrows. The path starts
at the source, first moves down along the y-axis in the direction
towards the destination node, then moves left along the x-axis, and
then moves forward towards the final destination. This figure will
act as a running example. Since it is difficult to visualize all the
nodes in three dimensions simultaneously, we only represent some of
the nodes here. The illustrated nodes will be relevant as we continue
the running example in Figure \ref{fig:One-possible-path}.}\label{fig:threeDimensionalupdownleftrightbackforth}

\end{figure}

For three dimensions, we use a natural generalization of the up-down
left-right protocol called the \emph{up-down left-right back-forth
protocol}. For discussion, we define the
up-down direction as the $y$-axis, left-right as the $x$-axis, and
the back-forth as the $z$-axis. See Figure \ref{fig:threeDimensionalupdownleftrightbackforth}
for an example of this labelling, as well as a particular path followed
by one message during the execution of a full communication step of
this protocol. 

This protocol is a natural and straightforward extension of the up-down
left-right protocol used in two dimensions. To determine in which
direction a node should send a message (up, down, left, right, backward,
or forward), the processor node looks at the destination address.
If the destination node is not on the same y-coordinate, up-down direction,
this node shall send the message parallel to this axis in a direction
closer to the destination node. If the destination is on the same
y-coordinate, then the node shall look to the x-coordinate, and send
left-right along this direction closer to the destination. If the
message is already on the same y- and x-coordinates, then the node
shall send along the z-axis, back-forth direction, in a direction
closer to the destination. Finally, if the node is the intended destination
for the current message, that node shall keep the message in its memory,
along with all other received messages, until the next computation
step.

Since each message has an address of the intended destination, the
messages are of size $\log(N)$. Note that once an orientation is
chosen for the nodes with a particular communication graph, the up-down
left-right back-forth protocol can be carried out, so long as there
are no conflicts bigger than $O(\log(N))$.

\subsection{Mead architecture}

In \cite{meadMemoryCost}, the Mead memory architecture is introduced,
and it is shown that a memory access takes a number of clock cycles
that is logarithmic in the size of the memory. Since we will be storing
addresses of size $\log(N)$ in our memory, for convenience we can
also assume that each word size is $O(\log(N))$. Since we assume
the memory of each node is of size $O(\log^{2}(N))$, each memory
access takes a number of clock cycles $O(\log(\log^{2}N))=O(\log\log(N))$.

\subsection{Detailed Passing Rules}

We must give a few extra details of our message-passing protocols
to ensure that the architecture can manage all these steps. For now,
we assume that there is one memory in each processor node, connected
to a single processor. This means a memory read or a memory write
can only happen once per processor instruction. This results in a
slight problem: if all messages are to be passed in parallel in the
up-down left-right protocol, how do we ensure that a processor is
not simultaneously required to both read and write to multiple memory
locations simultaneously? Since we are only concerned with asymptotic
scaling rules, we propose one particular way to deal with this. In
practice, there should be many methods to deal with this that can
save many constant factors. Our goal is only to show that scaling
rules hold. We describe our rules in two dimensions first. The three-dimensional
generalization is straightforward and we do not include the details
here.

We shall assume that there are nine memory stacks that are implemented
by the processor, the outgoing up, down, left, and right memory stacks,
the incoming up, down, left, and right stacks, and the final-destination
stack. We assume each of these stack locations has $\Theta(\log(N))$
memory locations (each of size $O(\log(N))$, enough to hold an address
and a message). Since there are nine of these, they can all fit in
an $O(\log^{2}N)$ sized memory. 

When a message is communicated from an adjacent node, the processor
computes the next direction in which the message should be sent, and
then the message is inserted into the incoming memory stack for that
direction. In the next one-hop step, the incoming memory stack is
treated as the outgoing memory stack, and the previous outgoing memory
stack pointer is re-initialized.

We now need to ensure that at each passing step, a processor must
only deal with one memory access at a time. There are two methods
we propose to deal with this. The first rule is the up-down-left-right
order rule. Every one-hop step should be divided into these four substeps.
We assume that each of these steps has at most $O(\log(N))$ substeps,
one for each possible message that needs to be sent. The rule is that
nodes should send first all of their up-going messages, then their
down-going messages, then their left-going and right-going messages,
respectively. This way, the receiving block knows which of its four
registers to check depending on which step it is currently in. Also,
it ensures that there is only one message to be processed by every
node during each sub-step. 

The second rule is the \emph{alternating checkerboard} rule. Consider
the nodes of the mesh network as coloured black and white like a checkerboard.
We divide passing steps into \emph{black passing steps} and \emph{white
passing steps. }The nodes are to alternate between white-passing steps
and black passing steps. During a black-passing step, each black node
should send at most one message (either an up, down, left, or right
message, depending on which up-down-left-right substep we are in).
Note that sometimes some nodes will send no messages because they
may have already sent all of them. During a white passing step, all
the white nodes do the same thing. This ensures that at each sub-step
of the one-hop step, a node must either process an outgoing message
or process an incoming message. Hence, memory does not need to be
accessed multiple times per step. This will multiply the number of
clock cycles in the one-hop step by a factor of two, and so our scaling
rule analyses apply.

\section{Main Theorem Outline}\label{sec:mainTheoremOutline}

The two main theorems of this paper, corresponding to two- and three-dimensional
mesh networks, involve sequences of communication graphs with bounded
maximum node degree. We let $N$ be the number of nodes of each graph,
where $N$ approaches infinity. We give an outline of the theorem
statements below, along with a proof outline. These will be proven
later in the paper.

Our two dimensional Theorem \ref{thm:2dTheorem} states that, for
sufficiently large $N$, there exists an orientation of the graph
with bounded node degree on the two-dimensional mesh network such
that the up-down left-right protocol has no conflicts bigger than
size $O(\log N)$.\textbf{ }The three-dimensional theorem, Theorem
\ref{thm:3dTheorem}, states that for sufficiently large $N$, there
exists an orientation of the bounded node degree graph on the three-dimensional
mesh network such that the up-down left-right back-forth protocol
has no conflicts bigger than size $O(\log N)$.

\textbf{Proof outline: }The proofs of these theorems first use Lemma
\ref{lem:LemmaScalingLemma} which we prove below. This is a combinatorial
graph theory lemma. The lemma can be summarized as follows: For a
graph of size $N$, consider placing this graph into a set of $N$
slots. Consider two subsets of those slots, $A$ and $B$. We suppose
that the sizes of $A$ and $B$ are bounded by some weak conditions.
The lemma proves that if the graph is randomly placed into all the
slots, the probability that the number of edges connecting the nodes
in slots $A$ and $B$ exceeds $\Omega(\log(N))$ is small, and scales
superpolynomially to zero.

The second part of the proof considers a particular possible node
where there could be a large routing conflict. We divide the types
of routing conflicts into different cases. For each such case, we
show that for any one-hop step, the number of possible source nodes
and destination nodes that could result in a conflict at this routing
node is bounded. In fact, they satisfy the conditions of Lemma \ref{lem:LemmaScalingLemma}
mentioned in the paragraph above. Hence, the probability of a size
greater than $\Omega(\log N)$ conflict of each type scales superpolynomially
to zero. 

A union bound argument then shows that the probability of \emph{any
}large conflict at \emph{any }mesh node at \emph{any }one hop step
approaches $0$. This means there must exist an orientation of the
graph in which no such large conflicts arise.

The rest of the paper fills in the details of this proof.

\section{The Main Technical Result: The Random Graph Placement Lemma}\label{sec:mainTechnicalLemma}

In this section we will prove a foundational Lemma for our result.
For the lemma, we consider a sequence of graphs of increasing numbers
of vertices and constant maximum node degree. If we randomly place
the graphs on a set of slots, the probability that two subsets of
the slots are connected by a large number of edges can be bounded. 

\subsection{Lemma Statement}

\begin{figure}
\includegraphics[scale=0.5]{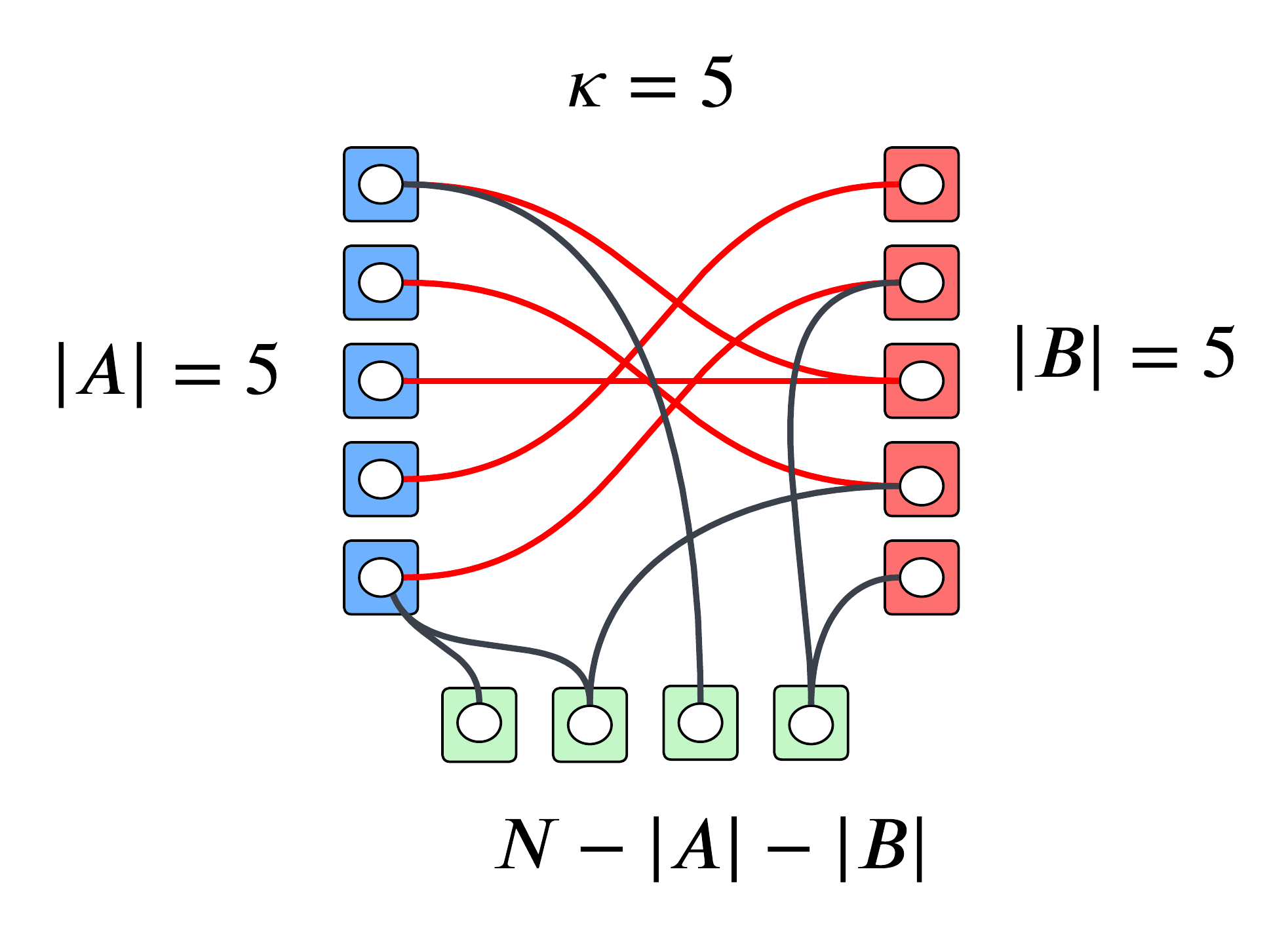}\caption{Diagram of $N=14$ slots (the shaded squares) with a particular orientation
of a graph. The special slots $A$ and $B$ are noted with their sizes
labelled and shaded in blue and red respectively. The remaining slots
are labelled with their size $N-|A|-|B|$ and shaded green. In this
case, there are $\kappa=5$ edges connecting the slots $A$ and $B$,
and thus we call it a $5$-connected configuration. For easy visibility,
the $5$ edges connecting nodes in the slots $A$ and $B$ are coloured
red. }\label{fig:slotsDiagram}
\end{figure}

We consider two sets of slots $A$ and $B$, as well as $N-|A|-|B|$
\emph{remaining }slots. Consider as well a graph $G=(V,E)$ with $N$
vertices in which the maximum degree of any vertex is $d_{\max}$.
Note there are $N!$ ways to permute the $N$ vertices of this graph
onto these (in total) $N$ slots. We call any such permutation a \emph{configuration.
}We show a diagram of a particular configuration of a graph on $N$
slots in Figure \ref{fig:slotsDiagram}.
\begin{lem}
\textbf{\label{lem:LemmaScalingLemma} }Let $|A|<c_{1}N^{\alpha}$
and $|B|<c_{2}N^{\beta}$, where $\alpha+\beta\le1$, and, as well,
$\alpha,\beta<1$, $\alpha,\beta>0$, and $c_{1},c_{2}>0.$ Let $R_{\kappa}$
be the event that a uniformly random configuration of the graph $G$
results in $\kappa>\log N$ edges connecting sets $A$ and $B$ when
these conditions are satisfied. When $N$ is sufficiently large, the
probability of this event is bounded by:
\[
P\left(R_{\kappa}\right)\le\exp(-\Omega(\log\log N\log N)).
\]

\end{lem}

The above Lemma we call the \emph{Graph Placement Lemma}.

\begin{IEEEproof}
(Of Lemma \ref{lem:LemmaScalingLemma}) Let $A$ and $B$ be disjoint
slot sets. We let $Z$ be the number of edges connecting the vertices
placed in $A$ and vertices placed in $B$.

Note that there are at most $d_{\max}|A|$ outgoing edges from the
slots in $A$, hence there are at most this many nodes that can be
selected for $B$. Let's fix a particular selection of $|A|$ nodes
of the graph and place them in the slots of $A$. Let $U$ be the
set of remaining vertices connected to the vertices selected for $A$.

Note that if there are $Z\ge\kappa$ edges connecting the nodes in
$A$ and $B$, then there are at least $r=\left\lceil \frac{\kappa}{d_{\max}}\right\rceil $
nodes of $U$ placed into $B$.

This implies that the event that there are $\kappa$ or more edges
connecting $A$ and $B$ is a subset of the event that there are $r$
vertices of $U$ chosen for $B$. Consider a particular subset of
$U$ of size $r$.

The probability that this particular subset of $U$ (of size $r$)
is placed in the $B$ slots, conditioned on the $|A|$ slots having
already been filled, is given by:

\[
\left(\frac{|B|}{N-|A|}\right)\left(\frac{|B|-1}{N-|A|-1}\right)\ldots\left(\frac{|B|-r+1}{N-|A|-r+1}\right)=\frac{|B|_{r}}{(N-|A|)_{r}},
\]
where in this case we define $X_{r}=X(X-1)\ldots(X-r+1)$.

To see this, consider the probability that the first such vertex is
selected. There are $|B|$ possible slots into which it can be placed,
out of $N-|A|$ remaining slots available. Conditioned on this being
selected, there are $|B|-1$ correct slots remaining, and $N-|A|-1$
remaining slots possible. This pattern continues until the last such
vertex is selected to be placed in the $|B|$ possible slots.

Thus, the probability that any subset of $U$ of size $r$ is chosen
can be upper-bounded by summing over all ${|U| \choose r}$ subsets
of $U$ of size $r$.

Hence, conditioned on a particular choice for the vertices in $A$:
\[
P(Z\ge\kappa)\le{|U| \choose r}\frac{|B|_{r}}{(N-|A|)_{r}}
\]

Now, for any choice of vertices to be placed in the slots of $A$,
the maximum size of the connected set $U$ is at most $d_{\max}|A|$,
because the maximum node degree is $d_{\max}$ and any connected vertex
must use up one of these edges. Hence:
\[
P(Z\ge\kappa)\le{d_{\max}|A| \choose r}\frac{|B|_{r}}{(N-|A|)_{r}}\le{d_{\max}|A| \choose r}\left(\frac{|B|}{N-|A|}\right)^{r}
\]

Now we use that
\[
{d_{\max}|A| \choose r}\le\left(\frac{ed_{\max}|A|}{r}\right)^{r}
\]
 to simplify the first term in the bound:
\[
P(Z\ge\kappa)\le\left(\frac{ed_{\max}|A||B|}{r\left(N-|A|\right)}\right)^{r}.
\]

Since we assumed $|A|<c_{1}N^{\alpha}$ and $\alpha<1$, we have that
for sufficiently large $N$, $|A|<\frac{N}{2}$. Also, by our assumption,
$|A||B|\le cN$. Thus, we conclude:
\[
P(Z\ge\kappa)\le\left(\frac{ed_{\max}(cN)}{r\left(N/2\right)}\right)^{r}=\left(\frac{2ed_{\max}c}{r}\right)^{r}=\left(\frac{C'}{r}\right)^{r}
\]
where $C'=2ed_{\max}c$, a constant greater than $0$. 

So now we let $\kappa=\log N$. Moreover, $r=\left\lceil \frac{\kappa}{d_{\max}}\right\rceil =\left\lceil \frac{\log N}{d_{\max}}\right\rceil $
implies $c'\log N-1\le r\le c'\log N+1$, where $c'=1/d_{\max}$.
This implies:
\[
P(Z\ge\kappa)\le\left(\frac{C'}{c'\log N-1}\right)^{c'\log N-1},
\]
and then exponentiating the logarithm of this expression and simplifying
gives us:
\[
P(Z\ge\kappa)\le\exp\left[\left(c'\log N-1\right)\log\left(\frac{C'}{c'\log N-1}\right)\right]=\exp(-\Omega(\log N\log\log N)).
\]
\end{IEEEproof}

\section{Two-Dimensional Message-passing}\label{sec:twoDimensionalMessagePassingResult}

\begin{figure}
\includegraphics[scale=0.5]{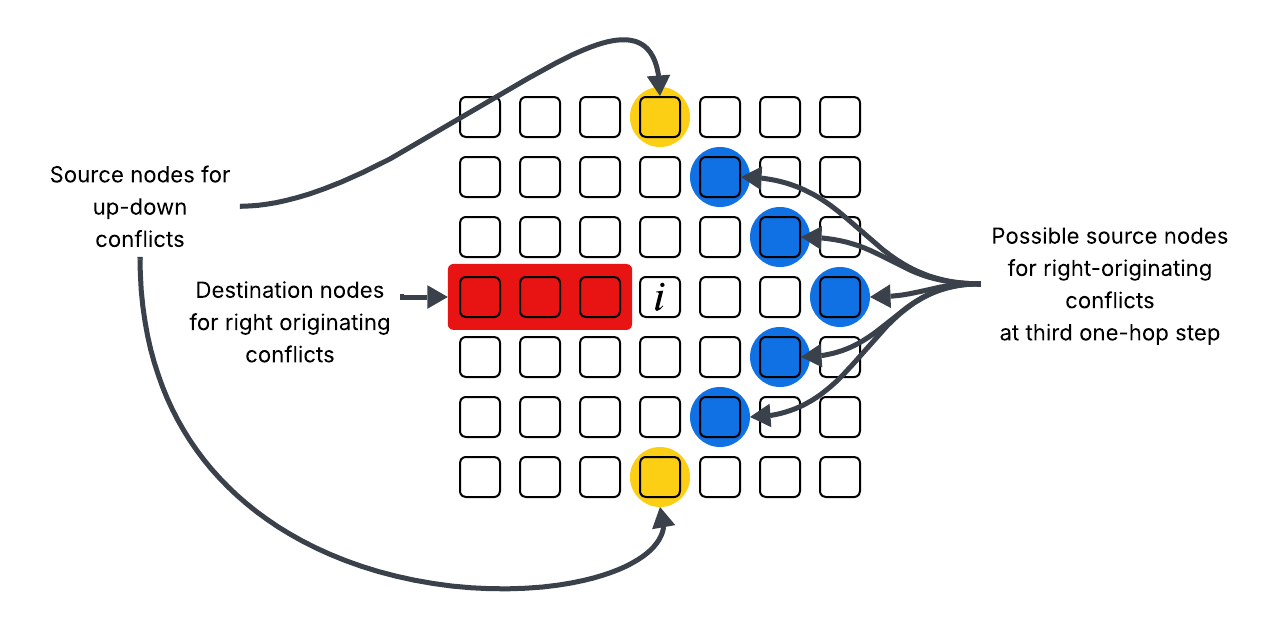}\caption{Figure of the source nodes (highlighted by blue circles), routing
node (labelled $i$), and locations of possible destination nodes
for messages resulting in right-originating conflicts after the third
one-hop step. The key idea of our proof is that the size of the set
of possible right-originating source nodes is at most $N^{\frac{1}{2}}$,
and the number of possible destination nodes is also at most $N^{\frac{1}{2}}$.
This is true at any one-hop step. We also label the location of source
nodes for possible up-down conflicts at the third one-hop step.}\label{fig:figureOfConflictSets}

\end{figure}

\subsection{Two-Dimensional Conflicts}

For this section, we consider two-dimensional mesh networks, and implement
the up-down left-right protocol as defined in Subsection (\ref{subsec:Two-Dimensions:-Up-down}).
For discussion, we define the up-down axis as the $y$-axis, and the
left-right axis as the $x$-axis. In Section \ref{sec:Three-Dimensional-Message-Passing},
we generalize this approach to three dimensions. Thus, this section
can be viewed as a warm-up. 

We now consider a particular mesh node, called the\emph{ routing node},
and a particular one-hop communication step $\tau$. We are concerned
with the number of messages being routed through this node at this
one-hop step. As we have discussed, the number of such messages being
passed is the size of the conflict at this node and one-hop step.
\begin{rem}
At first we consider only messages that are being routed \emph{through
}this particular routing node; we do not concern ourselves with messages
that reach a final destination at this node. We consider that later
when we put our analyses together.
\end{rem}
To simplify our analysis, we first consider right-originating conflicts.
These are messages that originate from the right side of the routing
node.

Each message routed through node $i$ at one-hop step $\tau$ must
have originated at a possible source node at Manhattan distance $\tau$
from the routing node\emph{. }Moreover, they can only lead to a specific
subset of mesh nodes, which we call the \emph{destination nodes. }Consult
Figure\emph{ }\ref{fig:figureOfConflictSets} for a labelling of possible
source and destination nodes that could result in a possible conflict
at routing node $i$. 

The rest of this section is devoted to bounding the size of the set
of possible source and destination nodes. Then, we will apply Lemma
\ref{lem:LemmaScalingLemma} (The Graph Placement Lemma) to show that
the probability that a random orientation of the graph onto the mesh
network results in a large conflict at routing node $i$ is small.
Finally, by using a union bound argument, we show that there exists
an orientation of the graph in which there are \emph{no }large conflicts.

\subsection{Bounding the size of up-down conflicts}\label{subsec:upDownConflictBound}

Observe Figure \ref{fig:figureOfConflictSets}. This shows the location
of the source nodes for potential up-down conflicts at the third one-hop
step. Note that there are only two such nodes (the one above, and
the one below). Hence, the size of an up-down conflict is at most
$2d_{\max}$. 

\subsection{Size of Possible Source and Destination Node Sets for Right-Originating
Conflicts}

Again, we are considering the number of messages that could be routed
through the $i$th mesh node during the $\tau th$ one-hop step.

Let us now consider those messages that are right-originating. Observe
from Figure\emph{ }\ref{fig:figureOfConflictSets} that the shape
of this set of possible source nodes is the half-diamond shape, and
such a set has size at most $\sqrt{N}$. 

Since we are using the up-down left-right protocol, once a message
has arrived at the routing node from a right-originating source node,
it can now only move in the left direction. Hence, the destination
node must be on the $x$-axis. The number of nodes on this axis is
at most $\sqrt{N}.$

\subsection{Bounding the probability of a size $\log N$ conflict or greater}

The two observations above lead us to our next lemma.
\begin{lem}
\label{Lemma:-RightOriginatingConflictLemma}The probability that
a random orientation of the graph on a mesh network (of sufficient
size to hold the graph) results in a right-originating conflict of
size greater than $O(\log N)$ at a node $i$ at a particular one
hop step scales as
\[
\exp(-\Omega(\log\log N\log N)).
\]
\end{lem}
\begin{IEEEproof}
Consider a particular node $i$ and one hop step $\tau$. By the arguments
in the preceding subsections, possible source nodes $B$ and destination
nodes $A$ have size bounded by $\sqrt{N}$. For any particular conflict
size $\kappa$ greater than $\log N$ to occur at this one-hop step,
there must be at least $\log N$ edges connecting $A$ and $B$. By
applying the Graph Placement Lemma \ref{lem:LemmaScalingLemma}, we
see that this occurs with probability bounded as $\exp(-\Omega(\log\log N\log N)).$
The result follows by the union bound over all conflict sizes from
$\log N$ to $d_{\max}\sqrt{N}$. 
\end{IEEEproof}

Using a symmetrical argument we also derive this lemma:
\begin{lem}
\label{Lemma:LeftOriginatingLemma}With the same setup as Lemma \ref{Lemma:-RightOriginatingConflictLemma},
the probability that a left originating conflict of size greater than
$\log N$ occurs at routing node $i$ at the $\tau$th one-hop step
is upper bounded by $\exp(-\Theta(\log\log N\log N)).$
\end{lem}

\subsection{Putting it together: There exists an orientation with no large conflicts}
\begin{lem}
The probability that a random orientation results in a conflict of
size greater than $2\log(N)+3d_{\max}$ at a particular routing node
$i$, at a one-hop step $\tau$ (an event we call $C_{i,\tau}$),
is bounded by  $\exp(-\Omega(\log(N)\log\log N))$.
\end{lem}
\begin{IEEEproof}
For this to occur, observe that at least one of the following must
occur:

1. A size $\log N$ or greater right-originating conflict called event
$C_{1}$.

2. A size $\log N$ or greater left-originating conflict called event
$C_{2}$.

3. A size greater than $2d_{\max}$ up-down conflict, called event
$C_{3}$.

4. Greater than $d_{\max}$ final messages have been received at node
$i.$

To prove this, suppose not. Suppose that none of these possibilities
is satisfied. Then, the total conflict size is at most $2\log(N)+3d_{\max}$.
The probability of any one of the events occurring is bounded by union
bound:
\begin{align*}
P(C_{i,\tau}) & \le\sum_{i=1}^{4}P(C_{i})\le2\exp(-\Omega(\log N\log\log N))+0+0\\
 & =\exp(-\Omega(\log(N)\log\log N)).
\end{align*}

For this inequality, we apply the bounds of Lemma \ref{Lemma:-RightOriginatingConflictLemma}
and Lemma \ref{Lemma:LeftOriginatingLemma} to bound events $1$ and
$2$. As well, the probability of event $C_{3}$ is $0$ because of
the argument in subsection \ref{subsec:upDownConflictBound}, and
$C_{4}$ is 0 because there are at most $d_{\max}$ edges leading
into any one vertex.
\end{IEEEproof}
\begin{lem}
The probability that a random configuration results in \emph{any }conflict
of size greater than $2\log(N)+3d_{\max}$ at any one-hop communication
step, at any node, is bounded by $\exp(-\Omega(\log N\log\log N))$.
\end{lem}
\begin{IEEEproof}
This flows from simply using union bound over the probability of this
event over all nodes of the mesh network (for which there are $N$),
and all possible one-hop step sizes $\tau$, each of which is at most
$2\sqrt{N}$. This is a sum over a polynomial number of cases, and
our bound on the probability for each of these cases scales at least
as fast as $\exp(-\Omega(\log N\log\log N))$, which scales superpolynomially
to $0$.
\end{IEEEproof}

\subsection{The Main Theorem for Two Dimensions: There exists an orientation
with no conflict bigger than $O(\log N)$}

We now state our first main Theorem:
\begin{thm}
\label{thm:2dTheorem}There exists an orientation of the nodes on
the two-dimensional mesh network in which the up-down left-right protocol
has at most a conflict of size $2\log(N)+3d_{\max}$.
\end{thm}
\begin{IEEEproof}
From the prior subsection, the probability of a large conflict existing
approaches $0$. Hence there must be at least one orientation with
no large conflicts. In fact, the probability that a randomly generated
orientation results in no large conflicts at any time is close to
$1$.
\end{IEEEproof}

\section{Three-dimensional Message-Passing}\label{sec:Three-Dimensional-Message-Passing}

In three dimensions, we consider the up-down left-right back-forth
protocol defined in Subsection \ref{subsec:3DUpdownLeftRightBackForthProtocolDefinition}.
We consider the probability that, using this protocol, a large conflict
occurs at a particular node at a particular one-hop step. We call
this node under consideration the \emph{routing node. }Without loss
of generality, let this node be located at $(0,0,0).$

There are two main types of conflicts that may occur. The first type
we call a \emph{through-conflict}, in which multiple messages are
passing through a routing node but have not yet reached their destination.
The second type we call a \emph{final-destination conflict}, in which
the message has reached its final destination at the routing node.

For each communication step of the computation, we consider the $\tau$th
one-hop communication step. Observe that if a message passes through
the routing node at one-hop step $\tau$, then at the step before
it, it was passed along one of the $x$, $y$, or $z$ axes. Alternatively,
the message has found its final destination at this particular node.
Hence, we divide the conflicts into the following types:
\begin{enumerate}
\item Right-originating left-right conflicts.
\item Left-originating left-right conflicts.
\item Back-originating back-forth conflicts.
\item Front originating back-forth conflicts.
\item Up-down conflicts.
\item Final destination conflicts.
\end{enumerate}
A conflict of size $\kappa$ of a particular type occurs at our routing
node if at this one-hop communication step there are $\kappa$ messages
that are to be held in its memory. For a message to arrive at our
routing node at exactly one-hop step $\tau$, it must have been sent
a Manhattan-distance $\tau$ starting from a possible \emph{source
node}. Moreover, there is a limited set of possible destination nodes
for such messages. For there to be a conflict of size $\kappa$ at
one-hop step $\tau$, there must be $\kappa$ edges connecting the
vertices in the possible source nodes to the possible destination
nodes.

It remains to simply bound the sizes of the possible source and destination
node sets and use Lemma \ref{lem:LemmaScalingLemma} to show that
large conflicts have low probability. That is the motivation of this
section. Then, using a straightforward union bound argument, we show
that the probability of a large conflict occurring anywhere at any
one-hop step must be small. 

In the end, this argument shows that there must be an orientation
of the graph on the mesh network that has only small conflicts across
all nodes and all one-hop steps.

\subsection{Up-down conflicts}\label{subsec:Up-down-conflicts-3Dbound}

For an up-down conflict to occur at the routing node at a particular
one-hop communication step, these messages must not have been sent
left-right or back-forth. Hence, they must originate on the same y-axis.
But for any one-hop communication step $\tau$, there are at most
only two source nodes that are this distance away (the one above and
the one below). Note that there may only be one or zero if $\tau$
exceeds the distance to the top or the bottom. 

The maximum node degree is $d_{\max},$ and hence the maximum size
of this type of conflict is $2d_{\max}$.

\subsection{Left-right conflicts}\label{subsec:Left-right-conflictsBound3D}

Left-right conflicts at a routing node $a$ occur because, at a particular
one-hop communication step, multiple messages are being sent to the
routing node along the $x$-axis. Each of these messages must have
originated at a source node at Manhattan-distance $\tau$ away from
the routing node. However, the set of possible destination nodes of
these messages need not be a fixed distance from the routing node.

We separately consider conflicts originating from the left and from
the right. Without loss of generality we consider the right-originating
conflicts. Let the set of possible destination nodes be $T_{z}$ and
the set of possible source nodes be $S_{r}$. We shall now bound the
size of each of these sets. 

\begin{figure}
\includegraphics[scale=0.3]{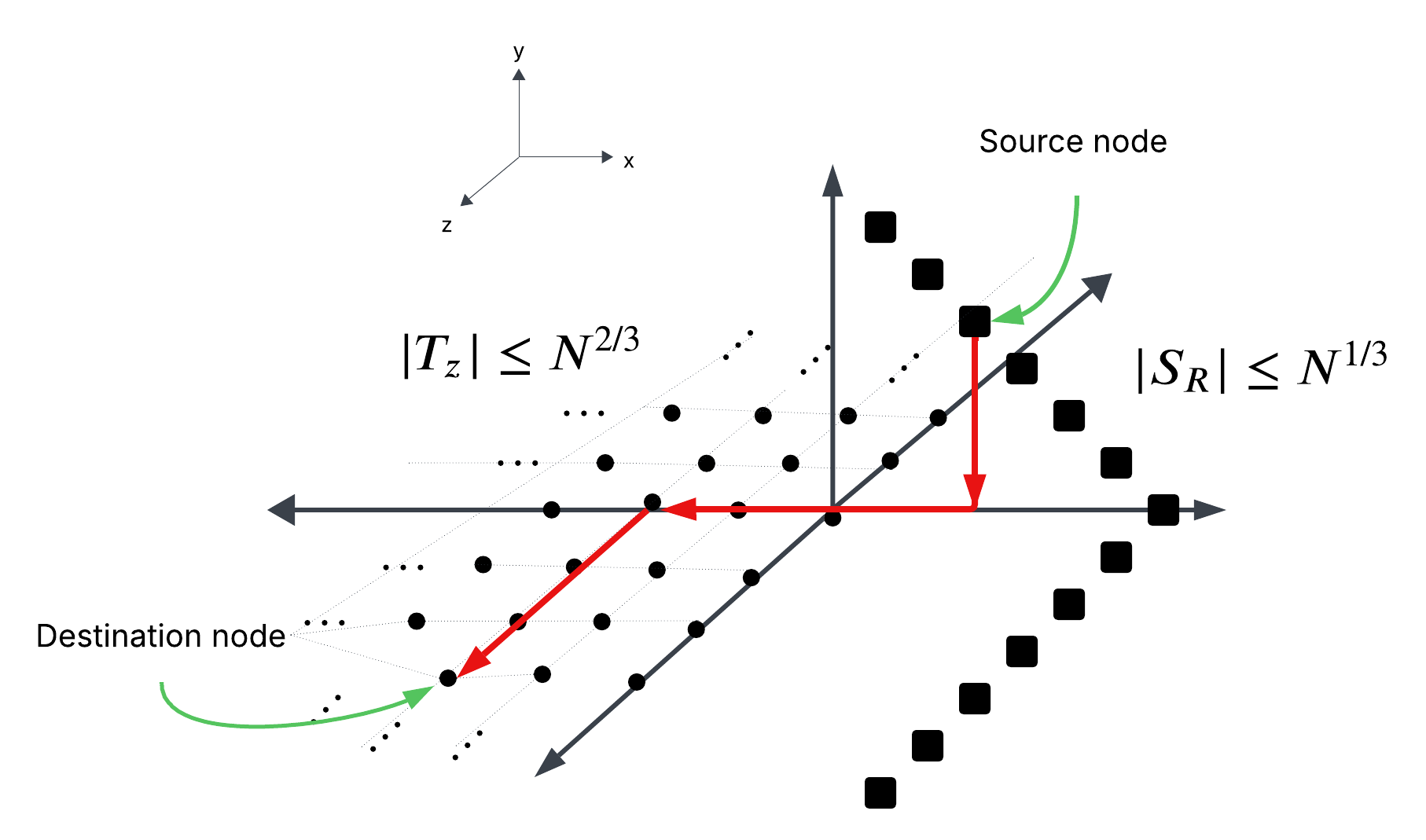}

\caption{Continuing with our running example, this figure shows the possible
right-originating source nodes (labelled $|S_{R}|$ and represented
as squares) whose messages would reach the routing node (located at
the origin for our example) after $7$ one-hop communication steps,
and also the possible destination nodes of these messages (labelled
$|T_{z}|$, and illustrated as dots). We also illustrate one possible
path of a message with the sequence of red arrows.}\label{fig:One-possible-path}

\end{figure}

\subsubsection{Question 1: How big is the possible source-node set?}

Recall we are labelling the routing node as the origin. We want to
bound the size of the set of the originating nodes.

Since messages pass through $(0,0,0)$ before going back-forth, the
originating nodes must be on the $z=0$ plane. See Figure \ref{fig:One-possible-path},
where we continue analyzing our running example for an illustration
of this case.

The source nodes must also be a distance $\tau$ away from this routing
node. There are at most $\sqrt[3]{N}$ possible such nodes originating
from the right. To see this, consider any height $y$. There is at
most $1$ node that is Manhattan distance $\tau$ away from the routing
node. Since there are at most $\sqrt[3]{N}$ possible heights for
a message to originate, there are at most $\sqrt[3]{N}$ possible
nodes distance $\tau$ from the routing node, for any $\tau$. 

\subsubsection{Question 2: How big is the possible destination-node set?}

The destination nodes must be in the $y=0$ plane, because the message
has already done the up-down step. This plane has at most $\sqrt[3]{N}\times\sqrt[3]{N}=N^{\frac{2}{3}}$
nodes in it.

\subsubsection{Probability of large conflict}

We can now bound the probability that a random orientation results
in a conflict of size $\kappa>\log N$ at our routing node. For such
a conflict to occur, there must be at least $\log N$ edges connecting
the source-node set $B$, which has size less than $N^{1/3}$, and
the destination set $A$, which has size less than $N^{\frac{2}{3}}$.
The exponents here add to $1$, and thus Lemma \ref{lem:LemmaScalingLemma}
applies. The probability of this is thus bounded by $P\left(R_{\kappa}\right)\le\exp(-\Omega(\log\log N\log N))$.

\subsection{Back-Forth conflicts}\label{subsec:Back-Forth-conflicts3Dbound}

We now consider back-forth conflicts. In a back-forth conflict at
one-hop communication step $\tau$, the messages have already potentially
moved in the up-down direction, and the left-right direction. Once
they reach the routing node, they only have movement along the $z$-axis
remaining. We divide such conflicts into two cases, the backward originating
and the forward originating back-forth conflicts. 

\subsubsection{Question 1: Backward originating back-forth conflicts: how big is
the originating set?}

\begin{figure}
\includegraphics[scale=0.3]{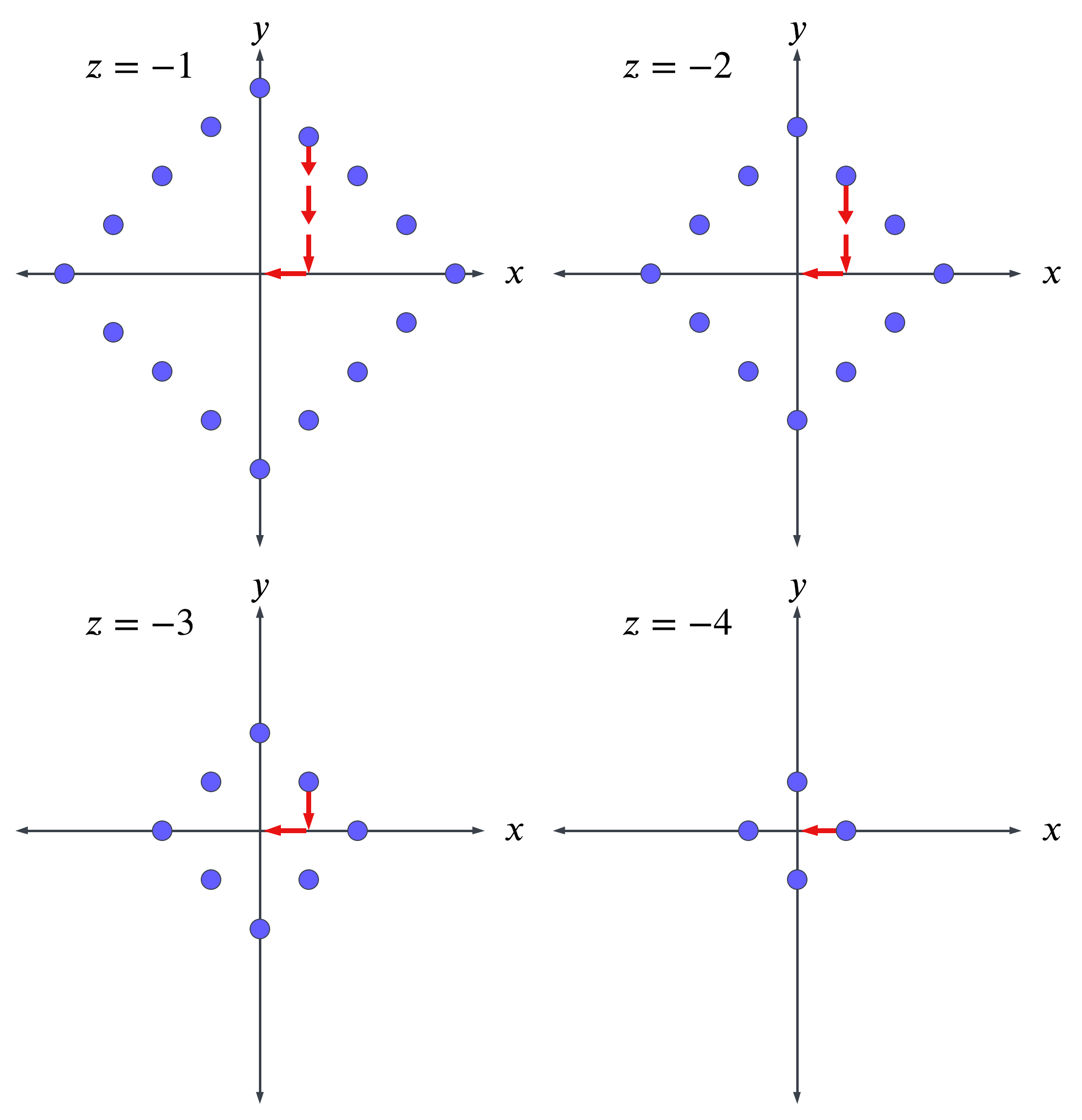}\caption{ Diagram of the possible source nodes for back-originating back-forth
conflicts at the $\tau=5$ one-hop step. Each diagram represents the
possible source nodes that are on the $z=-1,-2,-3$ and $-4$ planes.
On each diagram we label the number of steps it takes to reach the
$z$-axis for one representative node. So, for example, in the first
diagram (for $z=-1$), the nodes illustrated take $4$ one-hop steps
to reach the z-axis. From this point, each then has one more step
to reach the origin (which is the location of our routing node). Hence,
each of these nodes is $5$ one-hop steps away from the origin. For
the $z=-2$ diagram, each is $3$ one-hop steps from the $z$-axis,
and thus also $5$ one-hop steps from the origin. From this diagram
we can infer the very crude upper bound for the number of nodes at
a fixed distance $\tau$ from the origin at any particular plane where
$z$ is fixed. It can be at most $2$$N^{\frac{1}{3}}$, twice the
side length of our three-dimensional grid. Since $z$ can reach at
most to $-N^{1/3}$, we derive our $O(N^{2/3})$ bound on the size
of the possible source node set of the inequality in equation \ref{eq:backOriginatingBackForthConflictEquation}.}\label{fig:backForthConflictImage}

\end{figure}

We shall now derive a crude upper bound on the size of the set of
possible source nodes for the backward-originating conflicts. First,
let us consider possible source nodes that occur starting from the
$z=-1$ plane. For messages to be in conflict at our routing node,
they must be equidistant from the $z$-axis, because once they reach
the z-axis they have only one more hop to reach the routing node in
question. From inspection of the diagram in Figure \ref{fig:backForthConflictImage},
we can see that the number of such source nodes can be at most $2N^{\frac{1}{3}}$,
because that is twice the side length of the mesh network. In fact,
normally it would be much less than this, but for our crude upper
bound this is sufficient.

Similarly, for the $z=-2$ plane, there are also at most $2N^{1/3}$
such nodes. The reader should see that this crude upper bound should
hold for all possible $z$ coordinates.

The path that each message must follow must have length $\tau$. The
maximum length $\tau$ can be is $3N^{\frac{1}{3}}$ because this
is the maximum Manhattan distance between any two nodes in the mesh
network. Hence, we can bound the size of the source-node set $B$
as follows:
\begin{equation}
|B|\le\sum_{i=1}^{3N^{1/3}}2N^{\frac{1}{3}}\le6N^{\frac{2}{3}}\label{eq:backOriginatingBackForthConflictEquation}
\end{equation}

\subsubsection{Question 2: How big is the destination set?}

The destination set must occur only on the $z$-axis. Hence it is
of size at most $N^{1/3}$, because the number of nodes on the $z$-axis
is at most $N^{1/3}$.

\subsubsection{Probability of back-forth conflict of size $O(\log N)$ or greater}

Applying the Graph Placement Lemma \ref{lem:LemmaScalingLemma}, we
see that $|A|\le N^{1/3}$ and $|B|\le6N^{2/3}$, and hence our exponents
sum to $1$. Thus, we can apply our lemma and conclude that the probability
that a random permutation results in a back-originating back-forth
conflict of size $\text{\ensuremath{\kappa}>}\log N$ at any particular
one-hop step scales as $\exp(-\Omega(\log(N)\log\log(N)))$.

A symmetrical argument applies to front-originating conflicts.

\subsection{Summary}

From the arguments above, the probability of size $\log N$ conflicts
in each case is bounded by $\exp(-\Theta(\log N\log\log N))$.
\begin{thm}
The probability that a random orientation results in a conflict of
size greater than $4\log(N)+3d_{\max}$ at a particular routing node
$i$, at a one-hop step $\tau$ (an event we call $C_{i,\tau}$),
scales as $\exp(-\log(N)\log\log N)$.
\end{thm}
\begin{IEEEproof}
For this to occur, observe that at least one of the following must
occur:

1. A size $\log(N)$ or greater right-originating left-right conflict,
called event $C_{1}$.

2. A size $\log N$ or greater left-originating left-right conflict
called event $C_{2}$.

3. A size $\log(N)$ or greater back-forth conflict that is back originating,
called event $C_{3}$.

4. A size $\log(N)$ or greater back-forth conflict that is forward
originating, called event $C_{4}$.

5. A size greater than $2d_{\max}$ up-down conflict, called event
$C_{5}$.

6. More than $d_{\max}$ final messages have been received at node
$i.$

To prove this, suppose not. Suppose that none of these possibilities
is satisfied. Then, the total conflict size is at most $4\log(N)+3d_{\max}$.
The probability of this event is bounded by union bound.

We apply the bounds of Subsection \ref{subsec:Left-right-conflictsBound3D}
to bound events $C_{1}$ and $C_{2}$, and we apply the bounds of
Subsection \ref{subsec:Back-Forth-conflicts3Dbound} to bound events
$C_{3}$ and $C_{4}$. As well, the probability of event $C_{5}$
is $0$ because of the argument in Subsection \ref{subsec:Up-down-conflicts-3Dbound}
and $C_{6}$ is 0 because there are at most $d_{\max}$ edges leading
into one vertex. Hence, by union bound:
\begin{align*}
P(C_{i,\tau}) & \le\sum_{i=1}^{6}P(C_{i})\le4\exp(-\Omega(\log N\log\log N))+0+0\\
 & =\exp(-\Omega(\log N\log\log N)).
\end{align*}
\end{IEEEproof}
\begin{thm}
The probability that a random configuration results in \emph{any }conflict
of size greater than $4\log(N)+3d_{\max}$ at any one-hop communication
step, at any node, scales as $\exp(-\log N\log\log N)$.
\end{thm}
\begin{IEEEproof}
This flows from simply using union bound over the probability of this
event over all nodes of the mesh network (for which there are $N$),
all possible one-hop step indices $\tau$, each of which is at most
$3\sqrt[3]{N}$, and all potential conflict sizes, which range from
$\log N$ to no more than $Nd_{\max}$ (the maximum number of messages
that can be sent simultaneously). This is a sum over a polynomial
number of cases, and our bound on the probability for each of these
cases scales as $\exp(-\Omega(\log N\log\log N))$, which scales superpolynomially
to $0$.
\end{IEEEproof}
\begin{thm}
\textbf{\label{thm:3dTheorem} }There exists an orientation of the
nodes on the three-dimensional mesh network in which the up-down left-right
back-forth protocol has at most a conflict of size $4\log(N)+3d_{\max}$.
\end{thm}
\begin{IEEEproof}
The probability of a large conflict existing approaches $0$. Hence
there must be at least one orientation with no large conflicts. In
fact, the probability that a randomly generated orientation results
in no large conflicts at any time is close to $1$.
\end{IEEEproof}

\section{LDPC Scaling Rule}\label{sec:LDPCScalingSection}

We consider a sequence of LDPC codes implemented with Tanner graphs
in which the maximum node degree is bounded by a constant $d_{\max}$.
Message-passing algorithms for LDPC codes work by passing messages
sequentially between nodes of the Tanner graph. This algorithm can
be implemented in a natural way using the mesh network. 
\begin{rem}
We need to make one assumption about these upper bounds. We assume
that the messages passed in the LDPC decoding algorithm are represented
in at most $O(\log N)$ bits.
\end{rem}
For our message passing procedures, we require an address of size
$O(\log N)$ to be paired with each message. Thus, for our asymptotic
rules, allowing messages of size up to $O(\log N)$ does not change
our scaling results.
\begin{cor}
\label{cor:LDPCTannerGraphScalingRule}Each message-passing step of
the LDPC message-passing algorithm can be executed with area $O(N\log^{2}N),$
time that scales as $O(N^{1/2}\log(N)\log\log N)$, and thus energy
that scales as $O(N^{3/2}\log^{3}(N)\log\log N)$ in the two-dimensional
mesh network. In the three-dimensional network, if we accept the 3D
Circuit Folding Assumption, one message passing step can be executed
in time $T=O(N^{1/3}\log N\log\log N)$, and volume $V=O(N\log^{2}N)$,
and thus with energy that scales as $O(N^{4/3}\log^{3}N\log\log N)$.
\end{cor}
\begin{IEEEproof}
For two dimensions, let there be $N+K=O(N)$ computational nodes,
one for each node of the Tanner graph. Place the vertices of the Tanner
graph on the mesh network in a way that has at most $O(\log N)$ conflicts
for the up-down left-right protocol, which exists according to Theorem
\ref{thm:2dTheorem}. Implement the up-down left-right protocol to
execute the message passing steps of the decoding algorithm. Since
the maximum conflict size is at most $O(\log(N))$, and since each
message has size at most $O(\log(N))$, this will not overflow the
memory of any of the computational nodes (which have $\Theta(\log^{2}N)$
memory). The communication step can be done in at most $O(N^{1/2})$
one-hop steps, because the maximum number of hops between two nodes
in the two-dimensional mesh is $2N^{1/2}$.

We can divide the one-hop steps into the \emph{direction computation}
\emph{step}, where the direction to send each message is calculated,
and the\emph{ transmission-step, }where the messages are then sent
consecutively to the appropriate adjacent nodes. We introduce the
variables:

1. $n_{h}$: the number of one-hop steps.

2. $t_{d}$: the total number of clock cycles to compute the direction
to send all of the messages.

3. $t_{t}$: the number of clock cycles to transmit all the messages
to adjacent nodes.

4. $t_{h}=t_{d}+t_{t}$, the number of clock cycles to perform the
one hop step, which is equal to the time to compute the direction
to send the messages, plus the time to transmit the messages once
the direction is computed.

Thus, the total number of clock cycles to do the communication step
is $T\le n_{h}t_{h}$, the number of one-hop steps, times the time
per one-hop step.

For two dimensions, the direction-computation step takes $O(\log\log N)$
clock cycles for each message (each address must be passed through
the depth $O(\log\log(N))$ arithmetic logic unit). There are $O(\log N)$
such messages. Hence, direction-computation takes time $t_{d}=O(\log N\log\log N)$.
The transmission step takes time $t_{t}=O(\log(N))$ (one clock cycle
to pass each message to adjacent nodes for each size $O(\log(N))$
message). Hence, each one-hop step can be performed in time $t_{h}=O(\log N\log\log N+\log N)=O(\log N\log\log N$).
Since there are at most $O(N^{1/2})$ one-hop steps, the total time
of the communication step is at most $T=n_{h}t_{h}=O(N^{1/2}\log N\log\log N)$.

Since the switching activity factor is at most $1$ (all wires active
in each clock cycle), and the area of the mesh network is $A=\Theta(N\log^{2}N)$,
the energy of this computation is at most $AT=O((N\log^{2}N)(N^{1/2}\log N\log\log N))=O(N^{3/2}\log^{3}N\log\log N)$. 

For three dimensions, we apply our three-dimensional theorem (Theorem
\ref{thm:3dTheorem}). We place the $N+K$ computational nodes onto
the mesh network that results in at most $O(\log N)$ conflicts with
the up-down left-right back-forth protocol. Since the volume of each
node is $V=\Theta(\log^{2}N)$, this will mean that applying this
protocol for message passing will not overflow the memory of any node.
The number of one-hop steps is at most $n_{h}=O(N^{1/3})$.

The direction-computation step must take time $t_{d}=O(\log N\log\log(N)$)
(depth $\log\log(N)$ for each message, for $\log(N)$ messages).
Also, since the size of each face is $\Theta(\log^{\frac{4}{3}}N)$,
for the (at most) $O(\log N)$ messages (of size at most $O(\log N))$,
if one message is sent per clock cycle, the transmission step can
be done in $t_{t}=O(\log N)$ clock cycles.\footnote{Note that it may be possible to perform the transmission of messages
in less time than this, because we are not using the full face area
for transmission. As well, if the direction computation was parallelized,
this might result in an asymptotic saving for $t_{h}.$ However, this
would require some parallelization circuitry within each mesh node,
which we do not consider in this paper. } And so each one-hop step can be performed in time $t_{h}=O(\log N\log\log N+\log N)=O(\log N\log\log N).$ 

Hence, the number of clock cycles is at most $T=n_{h}t_{h}=O(N^{1/3}\log N\log\log N)$.
Since the volume of the mesh network is $V=N\log^{2}N$, and the switching
activity factor is at most $1$, the energy consumed during the message
passing step is at most $VT=O(N^{4/3}\log^{3}N\log\log N)$. 
\end{IEEEproof}
\begin{rem}
Note that the two-dimensional theorem reaches the almost-sure lower
bound of Theorem 2 in \cite{BlakeLDPCAlmostSure} up to a polylogarithmic
factor. This is a bound on almost all LDPC decoders generated according
to a uniform configuration distribution. We shall see how certain
LDPC codes can reach general lower bounds for \emph{any }coding algorithm
in the sections that follow.
\end{rem}
\begin{rem}
The above upper bound per iteration directly applies to the message
passing step of any other algorithm defined on a graph. Since LDPC
decoding is a prototypical example of an algorithm defined on a graph,
we present the theorem for this type of algorithm, but the upper bounds
are exactly the same for any algorithm defined on a graph with bounded
node degree. We give one other example of such an algorithm with sparse
matrix-vector multiplication in the next section.
\end{rem}

\section{Sparse Matrix-vector Multiplication}\label{sec:SparseMatrixVector}

Consider a sequence of sparse matrix-vector multiplications. Just
as in LDPC encoding, such a problem can be represented by a bipartite
graph (see Figure \ref{fig:matrixVectorGraphDiagram}). On the left
side of the graph, we can represent the input nodes. On the right
side, we represent the outputs of the matrix-vector multiplication.
In the analysis of matrix-vector multiplication, a key distinguishing
feature is whether the matrix itself is considered an input to the
problem. In our case, the matrix is \emph{not }an input, but rather
fixed before the computation takes place.

\subsubsection{Problem Definition}

The input to the computation is a vector of size $N$, where each
entry is chosen from a field $\mathcal{F}$. The size of the field
is $|\mathcal{F}|\le N$, and so the elements of the field can be
represented with $\log N$ bits. The matrix is known and fixed, where
the number of non-zero elements in each row and column is upper bounded
by $d_{\max}$. Moreover, the entries of the matrix are also from
the same field, and thus can be represented in $O(\log N)$ bits.
\begin{rem}
Alternatively, we can just think of the entries of the vector as floating
point or integer values represented with a fixed number of bits. The
key point is that for our upper bound, the number of bits to represent
the elements must be at most $O(\log N$).
\end{rem}
We define a sequence of matrix-vector multiplications as sparse if
the maximum number of non-zero entries in each row and column is $d_{\max}$. 

\begin{figure}
\includegraphics[scale=0.5]{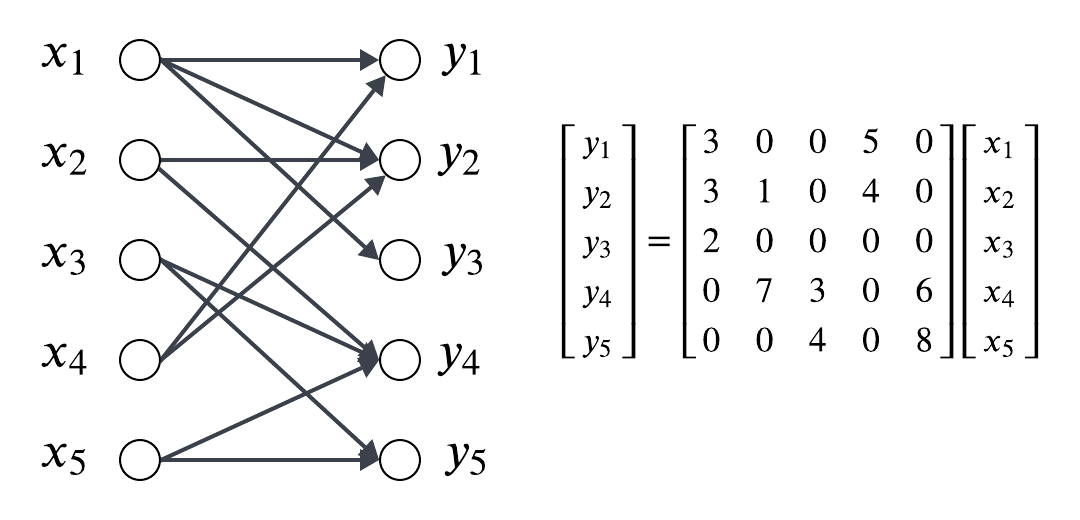}

\caption{Figure of the graph associated with a sparse matrix-vector multiplication,
along with its associated matrix. The nodes labelled with $x_{i}$
on the left represent the input vector values, and the nodes on the
right represent the outputs of the matrix-vector multiplication. To
illustrate what is happening here, consider the output node $y_{1}$;
it is connected to input nodes $x_{1}$ and $x_{4}$, since it must
calculate $3x_{1}+5x_{4}$, which corresponds to the matrix-vector
multiplication on the right.}\label{fig:matrixVectorGraphDiagram}
\end{figure}

In Figure (\ref{fig:matrixVectorGraphDiagram}) we show a natural
graph representation of an implementation of sparse matrix-vector
multiplication. In such an implementation, messages are passed from
input nodes to their corresponding output nodes. Since the number
of non-zero entries in each row and column is bounded, the vertex
degree of the graph representation is also bounded. Therefore, by
using the same argument as Corollary \ref{cor:LDPCTannerGraphScalingRule},
we can derive the following Corollary. Since the details are the same
as for the previous corollary, we omit the details of the proof:
\begin{cor}
\label{cor:matrixVectorMultiplicationUpperBound}(Two dimensions)
Sparse matrix-vector multiplication can be implemented on a two-dimensional
mesh network with area $A=O(N\log^{2}N)$ with a number of clock cycles
$O(N^{1/2}\log N\log\log N)$ and thus energy $O(N^{3/2}\log^{3}N\log\log N)$. 

(Three dimensions) Assuming the 3D Circuit-Folding Assumption, sparse
matrix-vector multiplication can be implemented on a three-dimensional
mesh network in volume $V=O(N\log^{2}(N))$, number of clock cycles
$O(N^{1/3}\log N\log\log N)$, and thus energy $O(N^{4/3}\log^{3}N\log\log N)$. 
\end{cor}
Note that this then implies that LDPC \emph{encoding }complexity,
which is just a matrix-vector multiplication, reaches the bounds above,
if the generator matrix is sparse.

\section{Using Parallelization to Construct Close to Energy and Time Optimal
$f(N)$-coding Schemes}\label{sec:parallelConstruction}

\subsection{The Fundamental Lower Bounds}

The results of \cite{blakePhDThesis,blakeKschischangEnergyLatencyReliabilityTradeoffs}
describe fundamental scaling rules for energy consumption of error
control coding circuits. Specifically, the authors define an $f(N)$-coding
scheme as a sequence of error control codes together with encoders
and decoders in which the rate stays above $0$ and the probability
of block error scales as $O(f(N))$. This must be associated with
a particular channel, which we assume to be a binary erasure channel
or a binary symmetric channel. By making hard decisions at the receiving
end, other channels like the additive white Gaussian noise channel
can be reduced to these channels, and thus are also subsumed in the
upper bound construction.
\begin{thm}
\textbf{\label{thm:FundamentalLowerBoundTheorem}The Fundamental Two-Dimensional
Lower Bounds (From Theorem 2 in \cite{blakeKschischangEnergyLatencyReliabilityTradeoffs}).
}All $f(N)$-coding schemes (for coding or decoding) have two-dimensional
energy consumption that scales at least as $\Omega(N\sqrt{-\log(f(N))})$.
For schemes that reach this lower bound on energy, the number of clock
cycles scales at least as $\Omega(\sqrt{-\log(f(N))})$.
\end{thm}
Note that when $f(N)$ decays exponentially, this lower bound reduces
to $E\ge\Omega(N^{3/2})$ and $T\ge\Omega(N^{1/2})$.
\begin{defn}
We call a coding scheme that reaches the energy lower bound up to
a polylogarithmic factor an \emph{energy-optimal coding scheme}. If
that coding scheme also reaches the clock cycle lower bound up to
a polylogarithmic factor, we call it an \emph{energy-time optimal
coding scheme}.

Note that it is possible to construct coding schemes with time complexity
better than the energy-time optimal lower bound; the problem is that
these come at a cost of more energy. Intuitively, what is happening
here is that more circuit area can be spent on creating greater wire
interconnectivity. Thus, more information can be sent per clock-cycle.
But, this requires potentially much greater circuit area, and thus
much greater energy. Energy-time optimal schemes are called this because
it is \emph{energy }that is optimized as a priority; and then for
this particular energy, we seek a scheme that has optimal time. An
example of a scheme that would take only a logarithmic number of iterations
would be the directly-implemented LDPC decoding scheme of \cite{BlakeLDPCAlmostSure},
but this would take $O(N^{2})$ area, and thus energy that scales
as $\Omega(N^{2}$), an $N^{1/2}$ factor away from optimal.

We also restate the three-dimensional lower bound for circuit energy.
Currently this lower bound does not have a similar bound on time that
exists in the two-dimensional case:
\end{defn}
\begin{thm}
\textbf{\label{thm:3dFundamentalLowerBoundtheorem}The Fundamental
Three-dimensional Lower Bounds (From Theorem 5 in \cite{blakeKschischangEnergyLatencyReliabilityTradeoffs}).
}All $f(N)$-coding schemes (for coding or decoding) have three-dimensional
energy consumption that scales at least as $\Omega\left(N\left(-\log f(N)\right){}^{1/3}\right)$.
\end{thm}

\subsection{The Upper Bound Construction}

Below, we define $\text{polylog}(N)$ to be any function that scales
polylogarithmically or more slowly in $N$. Note that it is not necessarily
the same polylogarithmic function in separate lines.

First, fix a communication channel with some channel statistic. Then,
suppose we have an $\exp(-cN)$-coding scheme for this channel for
some $c>0$ that has energy that scales as $O(N^{3/2}\text{polylog}(N))$
and time that scales as $O(N^{1/2}\text{polylog}(N))$. We shall use
this optimal scheme to show how to construct any energy and time optimal
$f(N)$-coding scheme for any $f(N)$ that scales to $0$ more slowly
than this exponential.

The following analysis works for both encoders and decoders in parallel.
Without loss of generality, we assume we are constructing decoders. 

Consider placing $N/M$ of the optimal exponential error circuits
in parallel, where each sub-circuit has block length $M$, thus creating
a circuit with total block length $N$.

We then simply choose $M$ to be a function of $N$ such that
\begin{equation}
\frac{1}{M}e^{-cM}=\frac{f(N)}{N}.\label{eq:equationForchoiceOfM}
\end{equation}

In particular, using the principal Lambert $W$ function, this can
be given by:
\[
M=\frac{1}{c}W_{0}\left(\frac{cN}{f(N)}\right).
\]

Note that the $W_{0}$ is defined as the function in which for positive
integer $x$, $W_{0}(xe^{x})=x$. However, we can simply use equation
\ref{eq:equationForchoiceOfM} for the rest of this derivation. We
also note that both the size of $M$ and number of sub-circuits must
be an integer. For ease of explanation we ignore this in our derivation,
as this does not change the scaling results.

First, observe that, from our assumption, $f(N)$ decays more slowly
than $e^{-cN}$, and so $f(N)\ge e^{-cN}$. If this is true, then
$M<N$. Hence, this is a valid choice for the size of each sub-circuit
(which, of course, cannot exceed $N$). To see this, if $M>N$ then
\[
f(N)=\frac{N}{M}e^{-cM}<e^{-cN}.
\]
An error occurs if and only if at least one of the decoders makes
an error, and so by union bound: 
\begin{equation}
P_{\mathrm{e}}\le\frac{N}{M}e^{-cM}=\frac{N}{N}f(N)=f(N).\label{eq:unionBoundEquationForParallelScheme}
\end{equation}
Hence, this is an $f(N)$-coding scheme. 

Since we assumed we started with an energy and time optimal scheme
with error probability that decays exponentially, then the energy
of each component in parallel scales as $O(M\sqrt{M}\text{polylog}(M))$.
Hence, the energy of such a parallel scheme is simply the sum over
all the individual energies of each of the individual sub-circuits
in parallel. Thus: 
\[
E\le\left(\frac{N}{M}\right)c'M^{3/2}\text{polylog}(M)=c'NM^{\frac{1}{2}}\text{polylog}(M)
\]
Note that if 
\[
\frac{1}{M}e^{-cM}=\frac{f(N)}{N}
\]
then:
\begin{align}
e^{-cM} & \ge\frac{f(N)}{N}\nonumber \\
-cM & \ge\log(f(N))-\log(N)\nonumber \\
M & \le\frac{-\log(f(N))}{c}+\frac{1}{c}\log N.\label{eq:boundOnM}
\end{align}
This implies:
\begin{align*}
E & \le N\sqrt{M}\text{polylog}(M)\\
 & \le N\sqrt{\frac{-\log(f(N))}{c}+\frac{1}{c}\log N}\text{polylog}(N)\\
 & \le N\left(\sqrt{\frac{-\log(f(N))}{c}}+\sqrt{\frac{1}{c}\log N}\right)\text{polylog}(N)\\
 & =O(N\sqrt{-\log f(N)}\mathrm{polylog(N))}.
\end{align*}
In the second line we use $M<N$, which implies $\text{polylog}(M)$
is bounded above by a suitable function that is polylogarithmic in
$N$. In the third line we use the inequality $\sqrt{x+y}\le\sqrt{x}+\sqrt{y}$.

Similarly, for bounding $T$, the number of clock cycles of all the
circuits in parallel is just the time of each circuit, which is $O(M^{\frac{1}{2}}\mathrm{polylog}(M))$.
Using a similar argument as above, as well as our bound from equation
\ref{eq:boundOnM}, we get:
\begin{align*}
T & \le O\left[\left(-\log(f(N))+\log N\right)^{1/2}\text{polylog}(N)\right]\\
\le & O\left[\left(\sqrt{-\log(f(N))}+\sqrt{\log(N)}\right)\text{polylog}(N)\right]\\
\le & O(\sqrt{-\log(f(N)})\text{polylog}(N)).
\end{align*}

This reaches the time lower bound of Theorem \ref{thm:FundamentalLowerBoundTheorem}
up to a polylogarithmic factor.

Also note that, if we want to avoid using the principal Lambert $W$
function, we can choose $M=-\frac{1}{c}\log\left(\frac{f(N)}{N}\right)$
for many choices of $f(N)$, and the reader should note that our upper
bounds still apply. This is only an invalid choice when $f(N)$ is
sufficiently close to $e^{-cN}$, which can make $M$ exceed $N$.
For this alternative choice of $M$, consider making a scheme with
error probability that scales as $e^{-N^{1/2}}$. Then we can choose:
\[
M=-\frac{1}{c}\log(e^{-N^{1/2}}/N)=\frac{1}{c}N^{1/2}+\frac{1}{c}\log N=\Theta(N^{1/2}).
\]

Thus, to create an energy-time optimal coding scheme, we need a primitive:
an $\exp(-\Theta(N))$-coding scheme that has energy that scales as
$O(N^{3/2}\text{polylog}(N))$ and time that scales as $O(N^{1/2}\text{polylog}(N))$.
We shall do this in Section \ref{sec:ExpanderGraphEnergyTimeOptimality}.

\subsection{Three Dimensional Parallelization Technique}

Using an analogous argument as above, we can also use a parallelization
technique to construct energy-optimal three-dimensional coding schemes.
We can also compute their time complexity (although we cannot compare
this to lower bounds since we lack three-dimensional time lower bounds
for energy optimal schemes). We first suppose we have an $\exp(-cN)$-coding
scheme with energy that scales as $E\le O(N^{4/3}\text{polylog}(N))$
and clock cycles $T\le O(N^{1/3}\text{polylog}(N))$. Then, by using
an analogous parallelization technique as above, for any $f(N)$ slower
than this $e^{-cN}$, we can construct an $f(N)$-coding scheme with
energy $E\le O\left[N\left(-\log(f(N))\right){}^{1/3}\text{polylog}(N)\right]$
and $T\le O\left[\left(-\log f(N)\right){}^{1/3}\text{polylog}(N)\right]$.

\section{Expander-graph Based LDPC Decoding Schemes are Energy-Time Optimal}\label{sec:ExpanderGraphEnergyTimeOptimality}

We now construct this energy-time optimal primitive using LDPC codes.

The existence of such LDPC decoders was conjectured in the thesis
\cite{blakePhDThesis}. However, the expander graph based construction
of Burshtein and Miller \cite{burshteinExpanderGraph} was overlooked
at the time. By proofreading an earlier draft of our paper using ChatGPT
Pro edition, we were pointed to this expander graph paper. After verifying
the applicability of these codes, we combine them with the argument
in the preceding section, to create the first known energy-time optimal
coding scheme. We describe an example construction in the subsections
below, which was also suggested by ChatGPT Pro.

\subsection{The Node Degree Distribution Construction}\label{subsec:NodeDegreeDistributionConstruction}

We restate two of the main results from \cite{burshteinExpanderGraph}.
Since we take this directly from the paper, refer to this paper for
more details of this construction. We will reproduce enough here to
derive our main corollary. 

First we consider a family of bipartite graphs defined by the following
sequences. First, let
\[
\boldsymbol{\lambda}=(\lambda_{l_{\min}},\ldots,\lambda_{l_{\max}})
\]
and
\[
\boldsymbol{\rho}=(\rho_{r_{\min}},\ldots,\rho_{r_{\max}})
\]
where $\lambda_{l}$ is the fraction of left nodes with degree $l$
and $\rho_{r}$ is the fraction of right nodes with degree $r$. Also,
we assume $l_{\min}$ and $r_{\min}$ are the minimal left and right
node degrees respectively. Similarly, $l_{\max}$ and $r_{\max}$
are the maximal left and right node degrees, respectively.

If there are $N$ left nodes and $R$ right nodes, then these sequences
can be used to define the number of left and right nodes of each degree.
The number of edges connected to the right nodes must equal the number
of edges connected to the left nodes. Hence, our quantities in $\boldsymbol{\lambda}$
and $\boldsymbol{\rho}$, as well as $R$ and $N$, must be chosen
such that:

\begin{equation}
E=\sum_{l=l_{\min}}^{l_{\max}}\lambda_{l}lN=\sum_{r=r_{\min}}^{r_{\max}}\rho_{r}rR.\label{eq:theConstraintForEdgeDegrees}
\end{equation}

We can now consider a random permutation of edges connecting the $E$
slots on the left with the $E$ slots on the right. 

This construction is associated with an LDPC code. The nodes on the
left are associated with codeword bits (and called the variable nodes)
and the nodes on the right are associated with the parity-check equations
(and called the check nodes). 

We also need to define an expander graph. To do so we borrow the notation
from \cite{burshteinExpanderGraph} in which for any set $V$ of variable
nodes, $\mathcal{N}(V)$ is the set of neighbors of $V$, and $\mathcal{E}(V)$
is the set of edges that connect $V$ with $\mathcal{N}(V)$.
\begin{defn}
(Directly from \cite{burshteinExpanderGraph}) A\emph{ $(\gamma,\delta,\boldsymbol{\rho},\boldsymbol{\lambda})$
expander }graph is a $(\boldsymbol{\rho},\boldsymbol{\lambda})$ bipartite
graph in which for any non-empty set $X$ of variable nodes with $|\mathcal{E}(X)|\le\gamma N$,
$|\mathcal{N}(X)|>\delta|\mathcal{E}(X)|$.
\end{defn}

\subsection{The Main Expander Graph Theorems}
\begin{thm}
\textbf{\label{thm:logarithmicIterationsAndExponentialErrorLemma}(Theorem
1 from \cite{burshteinExpanderGraph}) (Expander graph LDPC codes
have exponential error probability with a logarithmic number of iterations)}

Consider a $(\gamma,\delta,\boldsymbol{\rho},\boldsymbol{\lambda})$
expander. Define 
\[
\widetilde{\beta}=\beta+\frac{1-\beta}{l_{\min}},
\]
and assume 
\[
\beta\ge\frac{1}{2},\qquad\widetilde{\beta}<2\delta-1.
\]
Apply Gallager's hard-decision decoding algorithm with parameter $\beta$.

For an integer $N_{0}$ satisfying 
\begin{equation}
N_{0}\le\frac{\delta-\widetilde{\beta}}{(1-\widetilde{\beta})l_{\max}}\gamma N-\frac{\delta}{1-\widetilde{\beta}},\label{eq:numberOfErrorsCorrected}
\end{equation}
every pattern of $N_{0}$ errors in the received values is corrected
within 
\[
\left\lceil \log_{\frac{1-\widetilde{\beta}}{2(1-\delta)}}N_{0}\right\rceil +1
\]
decoding rounds. Here $\lceil x\rceil$ is the ceiling of $x$.
\end{thm}
\begin{lem}
\textbf{\label{lem:expanderGraphsExistLemma}(Lemma 1 from \cite{burshteinExpanderGraph})
(Expander graph codes of bounded maximum node degree exist, and occur
with high probability):} Consider a sequence of $(\boldsymbol{\rho,\lambda})$
bipartite graphs that are chosen at random uniformly as described
above in Section \ref{subsec:NodeDegreeDistributionConstruction}
in which the numbers of left nodes $N$ and right nodes $R$ grow
to infinity. Suppose $r_{\max}$ is bounded above and independent
of $N$. Also, suppose $l_{\min}$ is bounded below and independent
of $N$. Then, for any $\delta<1-1/l_{\min}$, there is a sufficiently
small $\gamma>0$ that is independent of $N$, such that with probability
$1-o(N)/N$, the randomly generated graph is a $(\gamma,\delta,\boldsymbol{\rho},\boldsymbol{\lambda})$
expander.
\end{lem}
Note: We make explicit that the sufficiently small $\gamma$ is independent
of $N$, which is not explicitly stated in the original Lemma statement.
This is apparent by inspection of the proof of Lemma 1, Appendix A,
page 17 of \cite{burshteinExpanderGraph}. Specifically, consider
the third equation after line 18 to see that $\gamma$ does not depend
on $N$. 

\subsection{The Expander Graph Theorems Imply $\exp(-\Omega(N))$ LDPC coding
Schemes Exist with Logarithmic Iterations}

We now proceed to the main derivation of this section, which is a
direct implication of the Expander Graph argument from \cite{burshteinExpanderGraph}.

First, let 
\[
l_{\min}=l_{\max}=6
\]
and 
\[
r_{\min}=r_{\max}=12.
\]

Then, the constraint of equation \ref{eq:theConstraintForEdgeDegrees}
implies that if there are $N$ variable nodes there are $N/2$ check
nodes. Hence there are at most $N/2$ independent rows of the parity-check
matrix, and the code has rate at least $1/2$. 

Applying Lemma \ref{lem:expanderGraphsExistLemma}, we can pick $\delta=4/5\le1-1/l_{\min}=5/6$.
Then, with probability that approaches $1$ as $N$ gets bigger, our
randomly produced graph is a $(\gamma,4/5,\boldsymbol{\rho},\boldsymbol{\lambda}$)
expander. Thus, for each $N$ in our sequence, choose such an expander
graph. 

Let the decoding parameter be $\beta=1/2$. Now, applying the parameters
of Theorem 1 of \cite{burshteinExpanderGraph} gives us 
\[
\widetilde{\beta}=\beta+\frac{1-\beta}{l_{\min}}=\frac{1}{2}+\frac{1-1/2}{6}=\frac{7}{12}.
\]
In this case, from the condition of Theorem \ref{thm:logarithmicIterationsAndExponentialErrorLemma},
we need $\widetilde{\beta}\le2\delta-1.$ We see that since $\delta=4/5$,
\[
\widetilde{\beta}=\frac{7}{12}<2\delta-1=2\left(\frac{4}{5}\right)-1=\frac{3}{5},
\]
and so the condition is satisfied.

Plugging these parameters directly into equation \ref{eq:numberOfErrorsCorrected}
implies that:

\[
N_{0}\le\frac{\delta-\widetilde{\beta}}{(1-\widetilde{\beta})l_{\max}}\gamma N-\frac{\delta}{1-\widetilde{\beta}}=\frac{4/5-7/12}{(1-7/12)(6)}\gamma N-\frac{4/5}{1-7/12}=\frac{13\gamma}{150}N-\frac{48}{25}.
\]

Note that the value $\gamma$ does not depend on $N$, and is some
small constant above $0$. 

Hence, this particular construction can correct a constant fraction
of errors in the received message. In particular, for sufficiently
large $N$, our construction can correct up to $\alpha N$ errors,
where $\alpha=\frac{13}{300}\gamma$. 

The number of iterations required to obtain this number of corrections
is logarithmic, with the logarithm having the following base:
\[
\frac{1-\widetilde{\beta}}{2(1-\delta)}=\frac{1-7/12}{2(1-4/5)}=\frac{25}{24}.
\]

This base is greater than $1$, and so the number of iterations scales
as $O(\log(N)/\log(25/24))=O(\log(N))$. 

Now we simply need to show that there is a channel in which this error
performance is exponential in block length $N$. 

Let $W_{N}$ be a random variable counting the number of errors in
our block length $N$ code. If we choose a channel in which $p<\frac{13}{300}\gamma$,
the probability of error is bounded by:

\begin{align*}
P_{\text{block }}(N) & \le P(W_{N}>\left\lfloor \alpha N\right\rfloor )\\
 & \le\exp(-ND(\alpha||p))
\end{align*}

where
\[
D(\alpha||p)=\alpha\ln\left(\frac{\alpha}{p}\right)+\left(1-\alpha\right)\ln\left(\frac{1-\alpha}{1-p}\right)
\]
This follows directly from the well known Chernoff bound for the tail
distribution of a binomial distribution. Hence, when $p<\alpha$,
we have an $\exp(-cN)$ coding scheme with a logarithmic number of
iterations.

\subsection{Tight Upper Bound Results}

We can now implement an LDPC decoder using the mesh network architecture
of Section \ref{sec:LDPCScalingSection}. Since that section describes
the complexity \emph{per iteration, }and there are a logarithmic number
of iterations required, this will only add a logarithmic factor to
the energy and time bounds. Hence, by using an appropriate expander-graph
based LDPC code and mesh network decoder, and requiring that the decoder
runs for only the appropriate logarithmic number of iterations:

\begin{thm}
There exists an $e^{-cN}$-coding scheme with area that scales as
$O(N\log^{2}N),$ time that scales as $O(N^{1/2}\log^{2}(N)\log\log N)$,
and thus energy that scales as $O(N^{3/2}\log^{4}(N)\log\log N)$
in the two-dimensional mesh network, for some $c>0$. Thus, for any
$f(N)$ that scales slower than $e^{-cN}$ to $0$, for some binary
symmetric channels, there exists an $f(N)$-coding scheme in which
decoding time $T\le O(\sqrt{-\log(f(N)})\text{polylog}(N))$ and $E\le O(N\sqrt{-\log(f(N))}\text{polylog}(N))$.
This reaches the fundamental lower bounds of Theorem \ref{thm:FundamentalLowerBoundTheorem}
up to polylogarithmic factors.
\end{thm}
We can also derive upper bounds for three-dimensional energy complexity,
and compare them to lower bounds on three-dimensional circuits from
\cite{blakeKschischangEnergyLatencyReliabilityTradeoffs}.
\begin{thm}
For some binary symmetric channels, there exists an $\exp(-\Theta(N))$-coding
scheme implementable in three dimensions with time that scales as
$O(N^{1/3}\log^{2}(N)\log\log N)$, and thus energy that scales as
$O(N^{4/3}\log^{4}(N)\log\log N)$. Thus, for any $f(N)$ that decays
more slowly than this, there exists an $f(N)$-coding scheme in which,
for decoding, 
\[
T\le O\left(\left(-\log(f(N))\right){}^{1/3}\text{\ensuremath{\mathrm{polylog}}}(N)\right)
\]
 and 
\[
E\le O\left(N\left(-\log(f(N))\right){}^{1/3}\text{\ensuremath{\mathrm{polylog}}}(N)\right).
\]
This reaches the fundamental energy lower bound on any three-dimensional
$f(N)$-coding scheme of Theorem \ref{thm:3dFundamentalLowerBoundtheorem}
up to polylogarithmic factors.
\end{thm}

\subsection{Limitations of this result}

We do not show that our construction of energy- and time-optimal $f(N)$-coding
schemes also reaches capacity. Moreover, we do not show that such
schemes exist for \emph{all }binary symmetric channels, but rather
only for ones for which we can apply the expander-graph approach.
These all remain significant open questions. Also, since generator
matrices are not necessarily sparse even when parity-check matrices
are sparse, we still have not shown a coding scheme which reaches
the complexity lower bounds for both encoding and decoding simultaneously.

\section{Conclusion}

In this paper we constructed tight (up to polylogarithmic factors)
upper bounds on the energy and time complexity of error control decoding
circuits. To do this, we derived general upper bounds for the time
and energy complexity of algorithms defined on communication graphs
of bounded vertex degree. Specifically, we considered these scaling
rule upper bounds for both two- and three-dimensional implementations
using the mesh network architecture. We highlighted the LDPC decoding
problem and the sparse matrix-vector multiplication problem as example
applications of this work. 

By applying the expander-graph based LDPC codes of \cite{burshteinExpanderGraph},
we showed that these LDPC codes can reach fundamental lower bounds
on time and energy in two dimensions for decoding derived in \cite{blakeKschischangEnergyLatencyReliabilityTradeoffs},
up to polylogarithmic factors. We also showed that these can reach
the fundamental lower bounds on energy for three dimensions.

There are many other open questions on this topic. We list a few below.
\begin{itemize}
\item The work in \cite{BlakeLDPCAlmostSure} shows that almost all Tanner
graphs generated according to a uniform configuration distribution
have high minimum bisection width. This implies lower bounds on the
energy of implementing LDPC decoding on the two-dimensional mesh network.
However, we conjecture that this \emph{also }implies comparable lower
bounds for three-dimensional circuits, but this has never been proven.
We suspect that lower bounds that closely match the upper bounds derived
herein are possible.
\item Our three-dimensional results rely on the 3D Circuit-Folding Assumption.
We assume that circuits implementable in two dimensional area $A$
can be folded into a three-dimensional cube of volume $V$ that scales
comparably to the original area $A$. A proof of this claim in general,
or even just for the specific circuits considered herein, is an area
of future work.
\item Understanding how graphs with \emph{unbounded }node degree, or node
degree that scales with the number of vertices $N$, can be routed
through mesh networks is also an open question.
\item Currently, the only lower-bound techniques we have for three-dimensional
circuits use the Grover information friction model. This approach
does not allow for simultaneous \emph{time }lower bounds. Hence, we
do not know if similar energy and time tradeoffs exist in three dimensions.
Nevertheless, we conjecture that the expander-graph LDPC construction
presented here is energy-time optimal in three dimensions. Specifically,
we suspect that any \emph{energy-optimal }$\exp(-cN)$-decoder in
three dimensions (which according to \cite{blakeKschischangEnergyLatencyReliabilityTradeoffs}
must have energy that scales as $\Omega(N^{4/3})$), must also require
clock cycles that scale as $\Omega(N^{1/3}$). We suspect similar
bounds would also apply for any $f(N)$-coding scheme. 
\item We have constructed energy-time optimal coding schemes herein. But
there is a more general question that still remains. If we allow for
greater energy consumption, the number of clock cycles can be reduced.
Thus, characterizing this tradeoff in general is an open question.
This tradeoff can be characterized in terms of lower bounds, as described
in \cite{blakeKschischangEnergyLatencyReliabilityTradeoffs}, but
the existence of achievable upper bound constructions for any $T(N)$
is an open question.
\item Our proposed implementation of the mesh network requires full processors
in each mesh node. Simple modifications could require only the direction-computation
circuitry. However, in either case, an \emph{entire address }of size
$\Theta(\log N)$ must be processed. This allows the mesh network
to be general purpose; in principle, the same architecture can carry
out any message-passing step defined by any graph with bounded vertex
degree. However, this might be overkill for \emph{specific }graphs.
Many graphs may have a specific structure that does not require message-passing
to use the full address of the destination node. Thus, for specific
graphs it may be possible to save some logarithmic factors in the
energy complexity by exploiting this special structure.
\item The two- and three-dimensional mesh networks described have the potential
to increase their processing capacity and, in the case of three dimensions,
their message-passing capacity. We presume that each mesh node has
only one processor, of size $O(\log N\log\log N)$, but the area/volume
of these nodes is $O(\log^{2}N)$, meaning that there can be more
processors per mesh node. This might allow for parallelization of
some of the message passing steps. Moreover, in the case of three
dimensions, each face has $O(\log^{4/3}N)$ surface area; however,
we only require that they pass $O(\log N)$ bits at a time. If the
direction computation steps and memory access steps could be parallelized,
this could potentially save some processing capacity. However, this
would require a memory management scheme so that memory accesses don't
dominate the complexity per message passed. Utilizing this latent
capacity remains an area of future work.
\end{itemize}

\section*{Acknowledgments}

We acknowledge that ChatGPT Pro Edition was used to verify mathematical
correctness of this paper. In the process, we were recommended a better
proof of the Graph Placement Lemma, as well as the tight upper bound
technique using expander-graph based LDPC codes. 
\renewcommand{\refname}{References}
\bibliographystyle{IEEEtran}
\bibliography{bibtextDoc}

@inproceedings{groverFundamental,
   author = "P. Grover and A. Goldsmith and A. Sahai",
   title = "Fundamental Limits on the Power Consumption of Encoding and Decoding",
   booktitle = "Proc. IEEE Int. Symp. Info. Theory",
   address={Cambridge, MA},
   year = 2012,
   pages = "2716--2720"
   }

@phdthesis{ThompsonThesis,
      author = {C. D. Thompson},
      title = {A Complexity Theory for {VLSI}},
      school = {Carnegie-Mellon},
      year = {1980},
      type = {Ph.{D}. Thesis},
    }

@ARTICLE{GroverInfoFriction, 
	author={Grover, P.}, 
	journal={IEEE Trans. Inf. Theory}, 
	title={Information Friction and Its Implications on Minimum Energy Required for Communication}, 
	year={2015}, 
	month={Feb}, 
	volume={61}, 
	number={2}, 
	pages={895-907}, 
}

@ARTICLE{meadMemoryCost,
author={Mead, C. and Rem, M.},
title={Cost and Performance of {VLSI} Computing Structures},
journal={IEEE Trans. Electron Devices},
volume={26},
number={4},
month={April},
year={1979},
pages={533-540},
}

@PHDTHESIS{HastadThesis,
	  author={J. Hastad},
	  title={Computational Limitations for Small Depth Circuits},
	  school={MIT},
	  year={1986},
	  }

@ARTICLE{BlakeLDPCAlmostSure,
  author={Blake, C. G. and Kschischang, F. R.},
  journal={IEEE Trans. Inf. Theory}, 
  title={On the {VLSI} Energy Complexity of {LDPC} Decoder Circuits}, 
  year={2017},
  volume={63},
  number={5},
  pages={2781-2795},
}

@ARTICLE{blakeKschischangEnergyLatencyReliabilityTradeoffs,
	author    = {Blake, C. G. and Kschischang, F. R.},
  	journal={IEEE Transactions on Information Theory}, 
  	title={Energy, Latency, and Reliability Tradeoffs in Coding Circuits}, 
  	year={2019},
  	volume={65},
  	number={2},
  	pages={935-946},
 }

@inproceedings{Thompson,
	   author = {C. D. Thompson},
	   title = {Area-time Complexity for {VLSI}},
	   booktitle = {Proc. ACM Symp. Theory of Comput.},
	   year = 1979,
	   pages = {81--88},
	   address={Atlanta, GA},
	   month=apr,
	   }

@book{bachmannNotation,
author={P. Bachmann},
title={Analytische Zahlentheorie},
publisher={ Teubner},
volume= 2, 
address={Leipzig, Germany},
year= 1894, 
}

@book{landauNotation,
author={ E. Landau},
title={Handbuch der Lehre von der Verteilung der Primzahlen},
publisher={ Teubner},
volume= 2, 
address={Leipzig, Germany},
year= 1909, 
}

@phdthesis{blakePhDThesis,
    author    = {Blake, C. G.}, 
    title     = {Energy Consumption of Error Control Coding Circuits},
    school    = {University of Toronto},
    year   = 2017,
    address   = {Toronto},
    month     = jun,
    url = {https://tspace.library.utoronto.ca/handle/1807/78035},
 type = {Ph.D. Thesis},
 }

@article{AblesonInformationTransferVLSIMultiplication,
  title = {Information transfer and area-time trade-offs for {VLSI} multiplication},
  author = {Abelson, H. and Andreae, P.},
  journal = {Communications of the ACM},
  volume = {23},
  number = {1},
  pages = {20--23},
  year = {1980},
  publisher = {Association for Computing Machinery},
}

@InProceedings{jinWireDelay,
  author =	{Jin, C. and Williams, R. R. and Young, N.},
  title =	{{A VLSI Circuit Model Accounting for Wire Delay}},
  booktitle =	{ITCS 2024},
  pages =	{66:1--66:22},
  ISBN =	{978-3-95977-309-6},
  ISSN =	{1868-8969},
  year =	{2024},
  volume =	{287},
}

@misc{gianinazzi2023spatialcomputermodelenergyefficient,
      title={The spatial computer: A model for energy-efficient parallel computation}, 
      author={Gianinazzi, L.  and Ben-Nun, T. and  Besta, M. and  Ashkboos, S. and  Baumann, Y. and  Luczynski, P. and  Hoefler, T.},
      year={2023},
      eprint={2205.04934},
      archivePrefix={arXiv},
      primaryClass={cs.DS},
      url={https://arxiv.org/abs/2205.04934}, 
}

@article{delYeyDeepLearningEstimatingEnergy,
title = {Estimating Deep Learning energy consumption based on model architecture and training environment},
journal = {Comp. Stand. Interfaces},
volume = {99},
pages = {104170},
year = {2027},
issn = {0920-5489},
doi = {https://doi.org/10.1016/j.csi.2026.104170},
url = {https://www.sciencedirect.com/science/article/pii/S0920548926000449},
author = { {del Rey}, S. and  Cruz, L. and Franch, X.  and Martínez-Fernández, S. },
}

@ARTICLE{dhananhayMonolithic3d,
  author={Dhananjay, K. and Shukla, P. and Pavlidis, V. F. and Coskun, A. and Salman, E.},
  journal={IEEE Trans. Circ. and Sys. II: Express Briefs}, 
  title={Monolithic 3{D} Integrated Circuits: Recent Trends and Future Prospects}, 
  year={2021},
  volume={68},
  number={3},
  pages={837-843},
}

@inproceedings{Ofman1962OnTA,
  title={On the Algorithmic Complexity of Discrete Functions},
  author={Ofman, Y.},
  year={1962},
booktitle={Dokl. Akad. Nauk SSSR},
volume={145},
number={1},
pages={48--51},
  url={https://api.semanticscholar.org/CorpusID:60961713}
}

@ARTICLE{burshteinExpanderGraph,
  author={Burshtein, D. and Miller, G.},
  journal={IEEE Trans. Inf. Theory}, 
  title={Expander graph arguments for message-passing algorithms}, 
  year={2001},
  volume={47},
  number={2},
  pages={782-790},
}

\end{document}